\documentclass[fleqn,usenatbib]{mnras}

\usepackage{newtxtext,newtxmath}

\usepackage[T1]{fontenc}

\DeclareRobustCommand{\VAN}[3]{#2}
\let\VANthebibliography\thebibliography
\def\thebibliography{\DeclareRobustCommand{\VAN}[3]{##3}\VANthebibliography}

\usepackage{graphicx}	
\usepackage{amsmath}	
\usepackage{csvsimple}  
\usepackage{dirtytalk}  
\hypersetup{
	colorlinks=true,       
	linkcolor=blue,        
	citecolor=blue,        
	filecolor=magenta,     
	urlcolor=blue         
}
\usepackage{booktabs}   

\usepackage{siunitx}
\DeclareSIUnit \h {\ensuremath{\mathit{h}}}
\DeclareSIUnit \pc {pc}

\title[Detection of weak lensing of SN Ia]{The Dark Energy Survey : Detection of weak lensing magnification of supernovae and constraints on dark matter haloes}

\author[DES Collaboration]{
\parbox{\textwidth}{
\Large
P.~Shah,$^{1}$
T.~M.~Davis,$^{2}$
D.~Bacon,$^{3}$
J.~Frieman,$^{4,5}$
L.~Galbany,$^{6,7}$
R.~Kessler,$^{8,5}$
O.~Lahav,$^{1}$
J.~Lee,$^{9}$
C.~Lidman,$^{10,11}$
R.~C.~Nichol,$^{3}$
M.~Sako,$^{9}$
D.~Scolnic,$^{47}$
M.~Sullivan,$^{12}$
M.~Vincenzi,$^{3,12}$
P.~Wiseman,$^{12}$
S.~Allam,$^{4}$
T.~M.~C.~Abbott,$^{13}$
M.~Aguena,$^{14}$
O.~Alves,$^{15}$
F.~Andrade-Oliveira,$^{15}$
J.~Annis,$^{4}$
K.~Bechtol,$^{16}$
E.~Bertin,$^{17,18}$
S.~Bocquet,$^{19}$
D.~Brooks,$^{1}$
D.~Brout,$^{46}$
A.~Carnero~Rosell,$^{20,14}$
J.~Carretero,$^{21}$
F.~J.~Castander,$^{6,7}$
L.~N.~da Costa,$^{14}$
M.~E.~S.~Pereira,$^{22}$
H.~T.~Diehl,$^{4}$
P.~Doel,$^{1}$
C.~Doux,$^{9,23}$
S.~Everett,$^{24}$
I.~Ferrero,$^{25}$
B.~Flaugher,$^{4}$
D.~Friedel,$^{26}$
M.~Gatti,$^{9}$
D.~Gruen,$^{19}$
R.~A.~Gruendl,$^{26,27}$
G.~Gutierrez,$^{4}$
S.~R.~Hinton,$^{2}$
D.~L.~Hollowood,$^{28}$
K.~Honscheid,$^{29,30}$
D.~Huterer,$^{15}$
D.~J.~James,$^{31}$
K.~Kuehn,$^{32,33}$
S.~Lee,$^{24}$
J.~L.~Marshall,$^{34}$
J. Mena-Fern{\'a}ndez,$^{35}$
R.~Miquel,$^{36,21}$
J.~Myles,$^{37}$
R.~L.~C.~Ogando,$^{38}$
A.~Palmese,$^{39}$
A.~Pieres,$^{14,38}$
A.~Roodman,$^{40,41}$
E.~Sanchez,$^{42}$
I.~Sevilla-Noarbe,$^{42}$
M.~Smith,$^{12}$
M.~Soares-Santos,$^{15}$
E.~Suchyta,$^{43}$
M.~E.~C.~Swanson,$^{26}$
G.~Tarle,$^{15}$
and N.~Weaverdyck$^{44,45}$
\begin{center} (DES Collaboration) \end{center}
}
\vspace{0.4cm}
\\
\parbox{\textwidth}{
$^{1}$ Department of Physics \& Astronomy, University College London, Gower Street, London, WC1E 6BT, UK\\
$^{2}$ School of Mathematics and Physics, University of Queensland, Brisbane, QLD 4072, Australia\\
$^{3}$ Institute of Cosmology and Gravitation, University of Portsmouth, Portsmouth, PO1 3FX, UK\\
$^{4}$ Fermi National Accelerator Laboratory, P. O. Box 500, Batavia, IL 60510, USA\\
$^{5}$ Kavli Institute for Cosmological Physics, University of Chicago, Chicago, IL 60637, USA\\
$^{6}$ Institut d'Estudis Espacials de Catalunya (IEEC), 08034 Barcelona, Spain\\
$^{7}$ Institute of Space Sciences (ICE, CSIC), Campus UAB, Carrer de Can Magrans, s/n, 08193 Barcelona, Spain\\
$^{8}$ Department of Astronomy and Astrophysics, University of Chicago, Chicago, IL 60637, USA\\
$^{9}$ Department of Physics and Astronomy, University of Pennsylvania, Philadelphia, PA 19104, USA\\
$^{10}$ Centre for Gravitational Astrophysics, College of Science, The Australian National University, ACT 2601, Australia\\
$^{11}$ The Research School of Astronomy and Astrophysics, Australian National University, ACT 2601, Australia\\
$^{12}$ School of Physics and Astronomy, University of Southampton, Southampton, SO17 1BJ, UK\\
$^{13}$ Cerro Tololo Inter-American Observatory, NSF's National Optical-Infrared Astronomy Research Laboratory, Casilla 603, La Serena, Chile\\
$^{14}$ Laborat\'orio Interinstitucional de e-Astronomia - LIneA, Rua Gal. Jos\'e Cristino 77, Rio de Janeiro, RJ - 20921-400, Brazil\\
$^{15}$ Department of Physics, University of Michigan, Ann Arbor, MI 48109, USA\\
$^{16}$ Physics Department, 2320 Chamberlin Hall, University of Wisconsin-Madison, 1150 University Avenue Madison, WI 53706-1390\\
$^{17}$ CNRS, UMR 7095, Institut d'Astrophysique de Paris, F-75014, Paris, France\\
$^{18}$ Sorbonne Universit\'es, UPMC Univ Paris 06, UMR 7095, Institut d'Astrophysique de Paris, F-75014, Paris, France\\
$^{19}$ University Observatory, Faculty of Physics, Ludwig-Maximilians-Universit\"at, Scheinerstr. 1, 81679 Munich, Germany\\
$^{20}$ Instituto de Astrofisica de Canarias, E-38205 La Laguna, Tenerife, Spain\\
$^{21}$ Institut de F\'{\i}sica d'Altes Energies (IFAE), The Barcelona Institute of Science and Technology, Campus UAB, 08193 Bellaterra (Barcelona) Spain\\
$^{22}$ Hamburger Sternwarte, Universit\"{a}t Hamburg, Gojenbergsweg 112, 21029 Hamburg, Germany\\
$^{23}$ Universit\'e Grenoble Alpes, CNRS, LPSC-IN2P3, 38000 Grenoble, France\\
$^{24}$ Jet Propulsion Laboratory, California Institute of Technology, 4800 Oak Grove Dr., Pasadena, CA 91109, USA\\
$^{25}$ Institute of Theoretical Astrophysics, University of Oslo. P.O. Box 1029 Blindern, NO-0315 Oslo, Norway\\
$^{26}$ Center for Astrophysical Surveys, National Center for Supercomputing Applications, 1205 West Clark St., Urbana, IL 61801, USA\\
$^{27}$ Department of Astronomy, University of Illinois at Urbana-Champaign, 1002 W. Green Street, Urbana, IL 61801, USA\\
$^{28}$ Santa Cruz Institute for Particle Physics, Santa Cruz, CA 95064, USA\\
$^{29}$ Center for Cosmology and Astro-Particle Physics, The Ohio State University, Columbus, OH 43210, USA\\
$^{30}$ Department of Physics, The Ohio State University, Columbus, OH 43210, USA\\
$^{31}$ Center for Astrophysics $\vert$ Harvard \& Smithsonian, 60 Garden Street, Cambridge, MA 02138, USA\\
$^{32}$ Australian Astronomical Optics, Macquarie University, North Ryde, NSW 2113, Australia\\
$^{33}$ Lowell Observatory, 1400 Mars Hill Rd, Flagstaff, AZ 86001, USA\\
$^{34}$ George P. and Cynthia Woods Mitchell Institute for Fundamental Physics and Astronomy, and Department of Physics and Astronomy, Texas A\&M University, College Station, TX 77843, USA\\
$^{35}$ LPSC Grenoble - 53, Avenue des Martyrs 38026 Grenoble, France\\
$^{36}$ Instituci\'o Catalana de Recerca i Estudis Avan\c{c}ats, E-08010 Barcelona, Spain\\
$^{37}$ Department of Astrophysical Sciences, Princeton University, Peyton Hall, Princeton, NJ 08544, USA\\
$^{38}$ Observat\'orio Nacional, Rua Gal. Jos\'e Cristino 77, Rio de Janeiro, RJ - 20921-400, Brazil\\
$^{39}$ Department of Physics, Carnegie Mellon University, Pittsburgh, Pennsylvania 15312, USA\\
$^{40}$ Kavli Institute for Particle Astrophysics \& Cosmology, P. O. Box 2450, Stanford University, Stanford, CA 94305, USA\\
$^{41}$ SLAC National Accelerator Laboratory, Menlo Park, CA 94025, USA\\
$^{42}$ Centro de Investigaciones Energ\'eticas, Medioambientales y Tecnol\'ogicas (CIEMAT), Madrid, Spain\\
$^{43}$ Computer Science and Mathematics Division, Oak Ridge National Laboratory, Oak Ridge, TN 37831\\
$^{44}$ Department of Astronomy, University of California, Berkeley, 501 Campbell Hall, Berkeley, CA 94720, USA\\
$^{45}$ Lawrence Berkeley National Laboratory, 1 Cyclotron Road, Berkeley, CA 94720, USA\\
$^{46}$ Department of Astronomy, 725 Commonwealth Avenue, Boston, MA 02215, USA\\
$^{47}$ Department of Physics, Duke University, 120 Science Drive, Durham, NC 27710\\
}
}
\date{Accepted XXX. Received YYY; in original form ZZZ}

\pubyear{2024}

\begin{document}
\label{firstpage}
\pagerange{\pageref{firstpage}--\pageref{lastpage}}
\maketitle

\begin{abstract}
The residuals of the distance moduli of Type Ia supernovae (SN Ia) relative to a Hubble diagram fit contain information about the inhomogeneity of the universe, due to weak lensing magnification by foreground matter. By correlating the residuals of the Dark Energy Survey Year 5 SN Ia sample (DES-SN5YR) with extra-galactic foregrounds from the DES Y3 Gold catalog, we detect the presence of lensing at $6.0 \sigma$ significance. This is the first detection with a significance level above $5\sigma$. Constraints on the effective mass-to-light ratios and radial profiles of dark-matter haloes surrounding individual galaxies are also obtained. We show that the scatter of SNe Ia around the Hubble diagram is reduced by modifying the standardisation of the distance moduli to include an easily calculable de-lensing (i.e., environmental) term. We use the de-lensed distance moduli to recompute cosmological parameters derived from SN Ia, finding in Flat $w$CDM a difference of $\Delta \Omega_{\rm M} = +0.036$ and $\Delta w = -0.056$ compared to the unmodified distance moduli, a change of $\sim 0.3\sigma$. We argue that our modelling of SN Ia lensing will lower systematics on future surveys with higher statistical power. We use the observed dispersion of lensing in DES-SN5YR to constrain $\sigma_8$, but caution that the fit is sensitive to uncertainties at small scales. Nevertheless, our detection of SN Ia lensing opens a new pathway to study matter inhomogeneity that complements galaxy-galaxy lensing surveys and has unrelated systematics.
\end{abstract}





\section{Introduction}
\label{sec:intro}

Type Ia supernovae (SNe Ia) magnitudes may be standardised using an empirical relationship derived from properties of their light curves and colours \citep{Phillips1993, Tripp1999} and an additional environmental adjustment which accounts for properties of the SN Ia host galaxy \citep{Kelly2010, Sullivan2010, Lampeitl2010}. After standardisation, the remaining intrinsic scatter (due to variation of the explosions) is approximately $\sigma_{\rm int} \sim ~0.1$ mag. Because of this low intrinsic scatter, SN Ia are ideal candidates to study gravitational lensing by intervening matter along the line of sight (LOS). 
\par
It is well known that the study of strongly-lensed variable point sources such as quasars and SNe Ia lead to constraints on distances in the universe by measurements of the time delay between multiple images \citep{Wong2020, Kelly2015, Rodney2021, Goobar2023}. However, each system requires detailed analysis and follow-up observations to constrain the foreground mass model, for which systematics larger than the statistical uncertainty appear to be present \citep{Birrer2020}. Additionally, strong lensing systems are rare, not straightforward to identify, and require extensive observation to constrain the relevant observables. In this paper, we take a different approach. We target the weak-lensing regime, in which only one image is seen and the magnification is at the percent level. We examine population-level statistics within the framework of a simple foreground mass model. Nevertheless, as we will show, we are still able to convincingly detect the presence of lensing and constrain our model parameters.
\par
In weak lensing we work to first order in the lensing convergence $\kappa$ (as defined below). SNe Ia are effectively point sources at cosmological distances, and their magnification $\Delta m_{\rm lens} \propto \kappa_{\rm F}$. By magnification, we mean relative to a homogeneous universe of the same average matter density (we have used the subscript F to refer to this as the \say{filled beam} convergence, again see below). Thus, SNe Ia seen along an overdense line of sight (LOS) will be brighter $(\Delta m_{\rm lens} < 0)$, and those along an underdense LOS (i.e., through voids) will be de-magnified $(\Delta m_{\rm lens} > 0)$. Gravitational lensing does not create or destroy photons, so it can be shown that $\langle \Delta m_{\rm lens} \rangle = 0+\mathcal{O}(\kappa^2)$ where the averaging is over a large number of SNe Ia (or any other type of) sources \citep{Kaiser2016}. The second-order effects are due to geometric corrections (the surface of constant redshift is no longer a sphere) and the non-linear conversion of fluxes to magnitudes, and will not be considered in this paper. 
\par
Weak gravitational lensing of an individual source reduces to the sum of many two-body \say{interactions}, which sample the distribution of matter along the line of sight. Therefore the dispersion between differing lines of sight, $\langle \Delta m^2_{\rm lens} \rangle = \sigma^2_{\rm lens}$, increases with distance to the source. As a rough guide we may expect $\sigma_{\rm lens} \sim 0.03$ mag for sources at $z = 0.5$, rising to  $\sigma_{\rm lens} \sim 0.08$ mag for sources at $z \sim 1.2$ \citep[][hereafter S22]{Shah2022}. Additionally, the probability distribution function (pdf) of $\Delta m_{\rm lens}$ along a randomly chosen LOS is highly skewed, with a small number of moderately magnified SN Ia balanced by a large number of weakly de-magnified ones. The simple reason for this is that a typical line of sight is more likely to pass through a large void than close to a halo \citep[see][for an estimation of skewness]{Kainulainen2011a}.
\par
Gravitational lensing impacts supernova cosmology in three ways. Firstly, it is one of the inputs to Malmquist bias calculations: at the magnitude limit of the survey, magnified SNe Ia scatter into the survey and de-magnified ones scatter out. The adjustment to SN Ia magnitudes due to this bias is computed using simulations, with a pre-specified lensing pdf and assuming the magnitude of the SN Ia is the sole determinant of selection \citep[see for example][]{Brout2019a}. Hence the assumed lensing pdf directly influences the corrected SN Ia magnitudes used to estimate cosmological parameters. A further issue is that given the potential for observational selection effects (such as a desire to avoid crowded foregrounds which might complicate spectroscopy), it is legitimate to ask whether a given SN Ia sample represents a fair sampling of the matter density of the universe: perhaps SN Ia datasets tend to preferentially select over- or under-dense lines of sight. 
\par
Secondly, lensing progressively increases the scatter of distant SNe Ia, decreasing their weighting in cosmological fits: proportionally more observations are thus needed to reach the same statistical precision at higher redshifts. In particular, SNe Ia at $z>0.6$ are useful to map the transition of the universe from deceleration to acceleration and confirm whether dark energy is indeed a cosmological constant or dynamical. If lensing can be estimated along a given line of sight, it can be treated as an additional standardisation parameter and corrected for. It was shown in S22 that de-lensing lowers scatter in Hubble diagram fits to the Pantheon sample \citep{Scolnic2018}. 
\par 
Thirdly, weak lensing is itself a source of cosmological information and may be used to determine parameters such as $S_8 = \sigma_8/\sqrt{\Omega_{\rm M}/0.3}$. Recently, a discrepancy has arisen between $S_8$ determined from the Cosmic Microwave Background (CMB), and from weak lensing as measured by cosmic shear. This discrepancy is persistent across multiple surveys covering different areas and using different analysis choices, and is moderately significant at the $\sim 2.5 \sigma$ level (see Fig. 4 of \citet{CosmologyIntertwined2022} for a summary of results). It has been proposed that this difference could be resolved by a late-universe suppression of the small-scale power spectrum at scales $k>0.1 h {\rm Mpc}^{-1}$, potentially due to increased baryonic effects \citep{Amon2022}. An alternative explanation may lie in the systematics of shear surveys such as intrinsic alignments. The magnification of SNe Ia offers a new way to probe the power spectrum of matter with unrelated systematics. 
\par
In S22, a forward model was proposed in which SN Ia lensing is assumed to be primarily due to dark matter haloes surrounding individual galaxies. This is justifiable, as the contribution to lensing from linear scale density fluctuations is expected to be small (this follows from Eqn. (5) of \citet{Frieman1996}, see also \citet{Kainulainen2011, Bahcall2014}). The model parameters are calibrated using SN Ia residuals to a Hubble diagram fit and photometric data of foreground galaxies.  
\par
In this paper, we have two main goals. Firstly, we use data from the Dark Energy Survey (DES) \citep{DES2016} and the model of S22 to calibrate our forward model of lensing. DES is well-suited to this purpose, because it combines a photometrically classified SN Ia survey with a galaxy survey conducted on the same platform. Confirming whether a supernova is Type Ia photometrically may reduce biases arising from spectroscopic selection preferring certain types of LOS (the host galaxy redshift is still confirmed spectroscopically). As the galaxy survey is conducted on the same platform, the foregrounds are effectively volume-limited (SNe Ia are somewhat fainter than a typical galaxy). Our primary goal is to detect the presence of lensing and determine features of the relationship between foregrounds and SN Ia magnitudes. Secondly, we de-lens the SN Ia magnitudes along their individual LOS and calculate the change in cosmological parameters, in order to determine if the DES SN Ia dataset is a fair representation of a homogeneous universe. As an application of our results, we calculate the observed dispersion of our lensing estimator and use it in a fitting formula given by \citet{Marra2013}, obtaining an estimate for $\sigma_8$. However, we note that the systematics of the fit are poorly understood for reasons we describe and view the result with skepticism pending further work. 
\par
Our paper is organised as follows. In Section \ref{sec:wl_by_gh}, we outline our modelling framework and likelihood. In Section \ref{sec:data}, we describe the data to be used. In Section \ref{sec:results}, we present our main results and discuss them in Section \ref{sec:summary}.

\section{Lensing model}
\label{sec:wl_by_gh}
In this section, we summarize the features of our model necessary to interpret the results. For more background and derivations, we refer the reader to the presentation in S22. 

\subsection{Weak lensing estimator $\Delta m$}
Working to first order in weak lensing convergence $\kappa$, the change in magnitude $m$ is $\Delta m  = -(5/\ln{10}) \kappa + \mathcal{O}(\kappa^2, \gamma^2)$ where $\gamma$ is the image shear. The zero point of $\kappa$ may be defined in two ways: relative to a homogeneous universe of uniform average matter density, denoted $\kappa_F$ or "filled-beam"; or relative to a zero matter density cylinder around the line of sight, denoted $\kappa_E$ or "empty-beam" \citep{Dyer1973}, with an unchanged background expansion. The choice does not matter for our results, but it is computationally convenient to work with the empty-beam definition. The two may be easily converted and we note $\kappa_F = \kappa_E - \langle \kappa_E \rangle + \mathcal{O}(\kappa^2)$. We may write it for supernova $i$ as the sum of the contribution from $N_i$ individual lenses along the LOS as

\begin{equation}
\label{eq:kappaprime}
    \kappa_{E,i} = \sum_{j=1}^{N_i} \kappa_{ij} \;.
\end{equation}
Adopting the lensing potential formalism of \citet{Schneider1985}, we have
\begin{equation}
\label{eq:conv}
    \kappa_{ij} = \frac{\Sigma_{ij}(\vec{\theta})}{\Sigma_{\rm c}} \; ,
\end{equation}
where the critical surface density $\Sigma_{\rm c}$ is
\begin{equation}
\label{eq:critsurfdens}
    \Sigma_{\rm c} = \frac{D_{\rm s}}{D_{\rm d} D_{\rm ds}} \frac{c^2}{4\pi G} \;\;,
\end{equation}
with $D_{\rm d}, \, D_{\rm s}, \, D_{\rm ds}$ the angular diameter distances to the lens, source and between lens and source respectively. The surface density $\Sigma$ is the integrated three-dimensional density $\rho$ of a given halo over the physical distance $l$ along the LOS specified by relative sky position $\vec{\theta}_{ij}$ between source $i$ and lens $j$ (adopting the Born approximation of an undeflected ray), and is 
\begin{equation}
\label{eq:surfdens}
    \Sigma_{ij}(\vec{\theta}) = \int \rho_{\rm halo}(\vec{\theta}_{ij}, l) dl \;.
\end{equation}
We take $\rho_{\rm halo}$ as a universal spherically-symmetric profile
\begin{equation}
    \label{eq:doublepowerlawhalo}
    \rho_{\rm halo} (r; \beta) = \frac{\delta_{\rm c} \rho_{\rm c}}{(\frac{r}{r_{\rm s}}) (1+(\frac{r}{r_{\rm s}}))^{\beta}} \; ,
\end{equation}
where $\rho_{\rm c} = 3 H(z)^2 / 8\pi G$ is the critical density of the universe at redshift $z$, $\delta_{\rm c}$ is a density parameter which can be calculated and $r_{\rm s}$ is the scale radius. Although halos are non-spherical, it has been shown that after averaging in a lensing calculation, spherical-symmetry is a very good approximation \citep{Mandelbaum2005}. $\beta$ defines the matter profile slope away from the core region and $\beta = 2$ reduces to the Navarro-Frenk-White (NFW) profile \citep{Navarro1997}. Analytical formulae may be derived for integer $\beta$ \citep[for the NFW case, see][]{Wright2000}, but Eqn. (\ref{eq:surfdens}) is straightforwardly computed numerically for general $\beta$.
\par 
The scale radius is defined as $r_{\rm s} = r_{200} / c$, where $r_{200}$ is radius where the fractional overdensity is $200$, and $c$ is the concentration parameter (not to be confused with the speed of light). In principle, $c$ should depend on halo mass, redshift and $\beta$. However, our data does not have much power to constrain $c$, as lines of sight do not pass sufficiently close to the cores of foreground galaxies to be influenced by it. Accordingly, we adopt the model of \citet{Mandelbaum2008} (which we refer to as M08) for $c(M_{200})$ as our fiducial choice, which has been calibrated using shear observations of galaxies from the SDSS survey. As the galaxies in SDSS have a lower average redshift than those of our sample, we also test the models of \citet{Duffy2008} (D08) and \citet{Cuartas2011} (C11) which have been calibrated against N-body simulations for $c(M_{200}, z)$. Finally, we will test consistency by letting $c$ vary as a free parameter. We will see in Section \ref{sec:results} below the specific concentration model adopted has no material effect on our results.

We define $r_{200}$ as a function of the mass $M_{200} = M(r<r_{200})$ it encloses via
\begin{equation}
\label{eq:m200}
    M_{200} = 200 \rho_{\rm c} \frac{4\pi}{3} r_{200}^{3} \; ,
\end{equation}
and then relate $M_{200}$ to the r-band galactic magnitude $M_{\lambda}$ by 
\begin{equation}
\label{eq:gammadefn}
    M_{200} = \Gamma \times 10^{0.4(M_{ \odot, \lambda} - M_{\lambda})} \; ,
\end{equation}
where $M_{ \odot, \lambda}$ is the solar absolute magnitude. We can expect that the mass-to-light ratio, $\Gamma$, depends in general on redshift, halo mass and morphology, and also absorbs residual Malmquist bias (discussed further below). In the simplest version of our model we take it to be constant; in this case it should therefore be seen as an \textit{effective} population average. We also test dependence on redshift and absolute magnitude, although in these cases our constraints are weaker.
\par
Our baseline model parameters are therefore $(\Gamma, \beta)$. The lensing estimate for a given SN Ia is then 
\begin{equation}
\Delta m_i = -(5/\log{10})(\kappa_{E, i} - \langle \kappa_{E}\rangle)
\end{equation}
where the average is the empty-beam convergence due to a homogeneous universe of physical matter density $\bar{\rho}$ from the observer to the source is
\begin{equation}
\langle \kappa_{\rm E} \rangle = \int_{0}^{z_s} \bar{\rho}(z)/\Sigma_c(z) dz \;\;. 
\end{equation}
We divide our SNe Ia into redshift bins, and set $\langle \kappa_{\rm E} \rangle = \sum_{z_i \in \rm bin_k} \kappa_{\rm E, i}$ so that $\langle \Delta m \rangle = 0 $ in each bin by construction.
\par
In summary, the key assumptions underlying our model are then :
\begin{itemize}
    \item weak lensing magnification is primarily due to haloes centred on galaxies,
    \item the halo density profile is statistically well approximated by a spherical universal halo profile,
    \item the lines of sight to SNe Ia are equivalent to a random sample of hosts,
    \item the masses of dark matter halos may be estimated from galactic magnitudes by a mass-to-light ratio.
\end{itemize}

\subsection{SN Ia distance residuals}
For our background cosmology, we assume a spatially-flat $\Lambda$CDM model (in \citet{SNKeyPaper} it was shown that more complex cosmologies are generally not preferred by SN data). The angular diameter distance $D_{\rm A}$ and luminosity distance $D_{\rm L}$ at late times are given by 
\begin{align}
\label{eq:cosmoformulae}
    D_{\rm L} (z) & =  \frac{c}{H_0} (1+z_{\rm obs}) \int_{0}^{z_{\rm cos}} \frac{dz'}{E(z')} \; , \\
    D_{\rm A} (z) & = D_{\rm L} / (1+z_{\rm obs})^2, \nonumber \\
    E(z) & = \sqrt{\Omega_{\rm M} (1+z_{\rm cos})^3 + 1-\Omega_{\rm M}}\; , \nonumber
\end{align}
where $H_0$ is the present day Hubble constant, $H(z) = H_0 E(z)$ and  $\Omega_{\rm M}$ is the present day matter density. $z_{\rm obs}$ refers to the observed heliocentric redshift, and $z_{\rm cos}$ the redshift corrected for peculiar velocities to the CMB rest-frame. When using standard candles, it is convenient to re-express the luminosity distance as the distance modulus
\begin{equation}
    \label{eq:mumodel}
    \mu_{\rm model} = 5 \, \mbox{log}_{10} (D_{\rm L}(z)/10 \mbox{pc}) \; .
\end{equation}
Our model for the matter density is 
\begin{equation}
\label{eq:rhomodel}
        \rho(\vec{r},z) = \rho_{\rm uniform}(z) + \sum \rho_{\rm halo}(\vec{r}_i,z)
\end{equation}
where $\rho_{\rm halo} (\vec{r}_i,z)$ is as defined in Equation \ref{eq:doublepowerlawhalo}. $\rho_{\rm uniform}(z)$ is a spatially uniform minimum density that is a function of redshift only; it represents the average remaining density of the universe if the virial masses of galactic halos were removed and is determined by the requirement that $\bar{\rho} = \rho_{\rm c}$. 
\par 
We determine the SN Ia distance modulus residuals $\mu_{\rm res}$ to the best-fit homogeneous cosmology Hubble diagram, obtained by minimizing 
\begin{equation}
\label{eq:chi2}
    \chi^2 = \mathbf{\mu}_{\rm res}^{T} \cdot \mathbf{C^{-1}} \cdot \mathbf{\mu}_{\rm res} \;\;,
\end{equation}
where $\mathbf{\mu_{res}} = \mathbf{\mu} - \mathbf{\mu}_{\rm model}$. $\mathbf{C}$ is the DES-SN5YR covariance matrix, which is the sum of statistical and systematic components \citep{Vincenzi2024}. $\mu$ is the apparent standardised (see Eqn. \ref{eq:trippnew} below) SN Ia distance modulus $\mu = m - M$, measured using scene modelled photometry and the B-band amplitude of a model template fitted by SALT3 \citep{Kenworthy2021}. We marginalise over the cosmological parameters $\Omega_{\rm M}$ and $H_0$ (which is degenerate with the fiducial SN Ia absolute magnitude $M$) but our results are largely unaffected by marginalisation (see for example Eqn. (\ref{eq:pearsoncorrelation}) below which is insensitive to the mean residual in each redshift bin).

\subsection{Lensing likelihood}
It is conventional in cosmological analyses to adopt a Gaussian likelihood as per Eqn. (\ref{eq:chi2}) for SN Ia residuals. This is not entirely accurate as in addition to potential intrinsic skew of SN Ia luminosities (due to variation in the physical conditions of the explosion), lensing introduces skew by moving the mode of the probability distribution to positive residuals above the Hubble diagram and adding a tail of magnified SN Ia below the Hubble diagram. However, as we will subtract our lensing estimate from the data vector below, we are justified in continuing to adopt the Gaussian form as will have removed this source of skew. We also note that as the photometric foreground redshifts $\mathbf{z}$ (expressed here as a vector over galaxies) on which we base our lensing estimate are uncertain, we must incorporate this into our likelihood. 
\par
For our likelihood $\mathcal{L}$ we write 
\begin{multline}
    -2 \log{\mathcal{L}}  = 
    (\boldsymbol{\mu}
_{\rm res} - \mathbf{\Delta m(\Gamma, \beta)})^{T} \cdot \mathbf{D^{-1}} \cdot (\boldsymbol{\mu}_{\rm res} - \mathbf{\Delta m(\Gamma, \beta))} \\
     + (\mathbf{z}- \mathbf{\bar{z}})^{T} \cdot \mathbf{P^{-1}} \cdot (\mathbf{z} - \mathbf{\bar{z}}) + \mathrm{const.}\;\;,
\end{multline}
where the first term on the r.h.s. is the likelihood of the residual $\boldsymbol{\mu}_{\rm res}$ adjusted for the lensing estimate $\Delta m(\Gamma, \beta)$, and $D = C - \mathrm{diag}(0.055 z_{i})$ is the DES-SN5YR covariance matrix amended to remove the added lensing uncertainty \citep[see][for a description of the how $C$ is determined]{Vincenzi2024}. The second term is the probability of the photometric redshifts given the true redshifts $\mathbf{\bar{z}}$. Approximating the redshifts as uncorrelated (there is little overlap between foregrounds), we may set $P$ to be the diagonal matrix $P_{ii} = \sigma_{z,i}$ and $0$ otherwise, where $\sigma_{z,i}$ is the redshift uncertainty output from the photo-$z$ algorithm. Assuming that the photo-$z$ errors are Gaussian distributed, we may marginalise over the unknown $\bar{z}$ \citep{Hadzhiyska2020} and obtain 
\begin{equation}
\label{eq:like1}
    -2 \log{\mathcal{L}} = (\boldsymbol{\mu}_{\rm res} - \mathbf{\Delta m(\Gamma, \beta)})^{T} \cdot \mathbf{C_{\rm lens}^{-1}} \cdot (\boldsymbol{\mu}_{\rm res} - \mathbf{\Delta m(\Gamma, \beta)}) + \mathrm{const.}\;\;,
\end{equation}
where $\mathbf{C}_{\rm lens} = D + APA^{T}$, and to first order
\begin{equation}
\label{eq:like2}
    A_{ij} = \frac{d \Delta m_{i}}{dz_{j}} \;\;,
\end{equation}
where $i$ is the index of the SN Ia, and $j$ the index of the foreground galaxy. As $A$ gives the response of the lensing estimate to the photometric redshift uncertainty, $C_{\rm lens}$ has the straightforward interpretation of being the original SN Ia covariance matrix $C$ with the lensing variance replaced by the uncertainty in the lensing estimator $\Delta m$ due to photometric redshifts. 
\par
While the term $APA^{T}$ may in principle be calculated (and it is equivalent to Eqn. (11) of \citet{Vincenzi2024}), it is convenient just to resample from the photometric redshift distribution and recalculate $\Delta m_i$. We generate 10,000 resamples and find photo-$z$ uncertainties contribute typically $< 1\%$ of the magnitude of the diagonal elements of $D$. This is small enough to justify our neglect of off-diagonal photo-z covariance.  
\par 
To determine if foregrounds and residuals are indeed connected by lensing, we calculate the bin-wise weighted linear Pearson correlation coefficient between $\Delta m_i$ and $\mu_{\rm res, i}$ as
\begin{equation}
\label{eq:pearsoncorrelation}
    \rho_k = \frac{\sum_{i} w_i (\mu_{i, \rm res} - \langle \mu_{\rm res} \rangle_w) \Delta m_i}{\sqrt{\sum w_i (\mu_{i, \rm res} - \langle \mu_{\rm res} \rangle_w)^2} \sqrt{\sum w_i \Delta m_i^2} } \;,
\end{equation}
where the weights $w_i = 1/{\rm C}_{{\rm lens},ii}$ and the averages are similarily weighted, the subscript $k$ refers to bin $k$, and the sum runs over all SN Ia in that bin. We adopt flat priors over the ranges $\Gamma \in (40,400)$ and $\beta \in (0.5, 4.0)$, and posteriors were computed using Polychord\footnote{\url{https://github.com/PolyChord/PolyChordLite}} \citep{Handley2015}.

\section{Data}
\label{sec:data}

\subsection{Supernovae}
We use the DES Y5 SN Ia data set as described in \cite{DataReleaseInPrep}. The SN Ia survey was conducted in 10 deep-field regions of the DES footprint. The survey has an average single visit depth of 24.5 r-band mag in fields X3 and C3, and 23.5 in the others. The SNe Ia redshifts range from $ 0.01 < z < 1.13$.  Supernova candidates are analysed using a machine-learning classifier whose input is the light curve shape, the output of which is the probability of being an SN Ia. The diagonal of the covariance is then adjusted for this probability, down-weighting likely contaminants but not discarding them altogether \citep{Vincenzi2021}. There are 1,829 SNe, and we exclude those with $z<0.2$ as the expected amount of lensing will be very low.
\par
The SN Ia host is set to be the source identified from co-added deep-field images \citep{Wiseman2020} that is closest in directional light radius to the SN Ia \citep{Sullivan2006, Gupta2016}. The redshift of the SN Ia is set to be the post-hoc measured spectroscopic redshift of the host galaxy, determined by the Australian Dark Energy Survey (OzDES) \citep{Lidman2020}. The possibility of host confusion (that is, an SN Ia may be allocated to the wrong galaxy and therefore given the wrong redshift) was analysed in \citet{Qu2023}, and the effect on the computed cosmology was found to be minimal. A potential complication in our analysis is that a misidentified host may mean that the true host is erroneously located in the foreground close to the line of sight and contributes a spuriously large amount to the lensing estimate. We discuss this further below. 

\subsection{Galaxies}
We use galaxies drawn from the Dark Energy Survey Y3 Gold Cosmology dataset \citep{Sevilla-Noarbe2021}, as the current public release of the deep-field catalog \citep{Hartley2022} only covers $\sim 30\%$ of the SN fields. Additionally, we wish to derive a calibration for our model parameters that can be used for lines of sight across the entire DES footprint, in order to facilitate future comparisons with shear studies. The Y3 Gold catalog is expected to be $90\%$ complete at $m_r = 23.0$ and the faintest sources categorised as galaxies are up to $m_r \sim 26$.
\par
Using the Y3 Gold flags as recommended in \citet{Sevilla-Noarbe2021} for extended objects, we select entries which are in an aperture of radius $8\arcmin$ around the line of sight to each supernova, with  $\texttt{FLAGS\_FOOTPRINT}=1$, $\texttt{EXTENDED\_CLASS\_MASH\_SOF}=3$, $\texttt{NEPOCHS\_R}>0$, $\texttt{FLAGS\_BADREGIONS}<4$, $\texttt{FLAGS\_GOLD}<8$ and $\texttt{SOF\_PSF\_MAG\_R}>17$. In aggregate, these flags select for high-confidence extended and extra-galactic objects and reduce contamination from artifacts, stars and photometric errors.
\par 
We do not exclude the region around the bright star $\alpha$ Phe as it is a large fraction of the E field, and for our purposes the foreground photometry of galaxies in that region is sufficiently accurate. For redshift $z=0.5$, $8\arcmin$ corresponds to a distance of $\sim 3$ Mpc. This is more than sufficient to capture the scales relevant to SN Ia lensing, and our results do not depend on aperture choice provided it is above $3\arcmin$. We use the survey-derived photometric redshifts $\texttt{DNF\_ZMEAN\_SOF}$ and discard galaxies that have unreliable photo-$z$ estimates (such as might arise from degeneracies in the photo-$z$ fitting process) determined as $\sigma_z /(1+z_p) > 0.2$, which reduces the number of our foreground galaxies by $\sim 8\%$. We find no fields that are masked to any significant degree in the foreground galaxy sample. After these cuts, our foreground sample consists of 804,484 galaxies or an average of $\sim 440$ per SN Ia.
\par
We must exclude the host galaxy --- if present in the Y3 GOLD catalog --- from our foregrounds by cross-matching the positions of the (deep-field) SN Ia hosts with the Y3 catalog using the criteria that the positions are within $4{\arcsec}$, and either of the DNF and BPZ photometic redshifts are compatible with the deep-field host at the $5\sigma$ level using the catalog redshift error.\footnote{Both DNF and BPZ galaxy redshifts are used, as we have found instances where the reported DNF error appears to be under-stated.} If the nearest Y3 Gold object does not fulfill these criteria, it is assumed to be a foreground. We remind the reader we always use the host spectroscopic redshift for the SN Ia. 
\par
 We have tested the robustness of our results by varying the aperture radius for the foregrounds between $1\arcmin - 8\arcmin$, the choice of concentration model (M08, D08 and C11 and fixed values of the concentration parameter $c$ from $5-13$), the photo-$z$ accuracy criterion from $0.1 - 0.8$ and the criteria for deciding whether the nearest galaxy in Y3 Gold is the host galaxy or a foreground from $3\sigma$ to $7\sigma$. We found the typical variation in our correlation result for these analysis choices to be small compared to the statistical error, and generally $<0.25\sigma$(stat) (see Section \ref{subsec:correlation} below). Accordingly, we judge that systematics that can be parametrically estimated are not significant to our results. 
\par 
We derive the absolute magnitude $M_{\lambda}$ of the galaxy in a given passband as 
\begin{equation}
    M_{\lambda} = m_{\lambda} - \mu(z_p) - K_{\lambda}  \;,
\end{equation}
where $m_{\lambda}$ is the apparent magnitude \texttt{SOF\_CM\_MAG\_CORRECTED} corrected for Milky Way extinction. The K-corrections $K_{\lambda}$ are computed to $z=0$ using $(griz)$ passbands and the software package \texttt{kcorrect v5.0}\footnote{\url{https://github.com/blanton144/kcorrect}} \citep{Blanton2007}. The distance modulus $\mu$ is derived using the photometric redshift $z_p$ by Equations (\ref{eq:cosmoformulae}, \ref{eq:mumodel}), with cosmological parameters from the supernovae fit. $z_p$ also determines the impact parameter $b = \theta D_d(z_p)$ using formulae (\ref{eq:cosmoformulae}), the critical surface density $\Sigma_c(z_{\rm SN}, z_p)$ using Equation (\ref{eq:critsurfdens}) with $\rho_{\rm c}(z) = 3 H(z)^2 / 8 \pi G$ which in turn determines the halo physical radius $r_{200}$ by Equation (\ref{eq:m200}). 
\par
\begin{figure}
    \centering
    \includegraphics[width=\columnwidth]{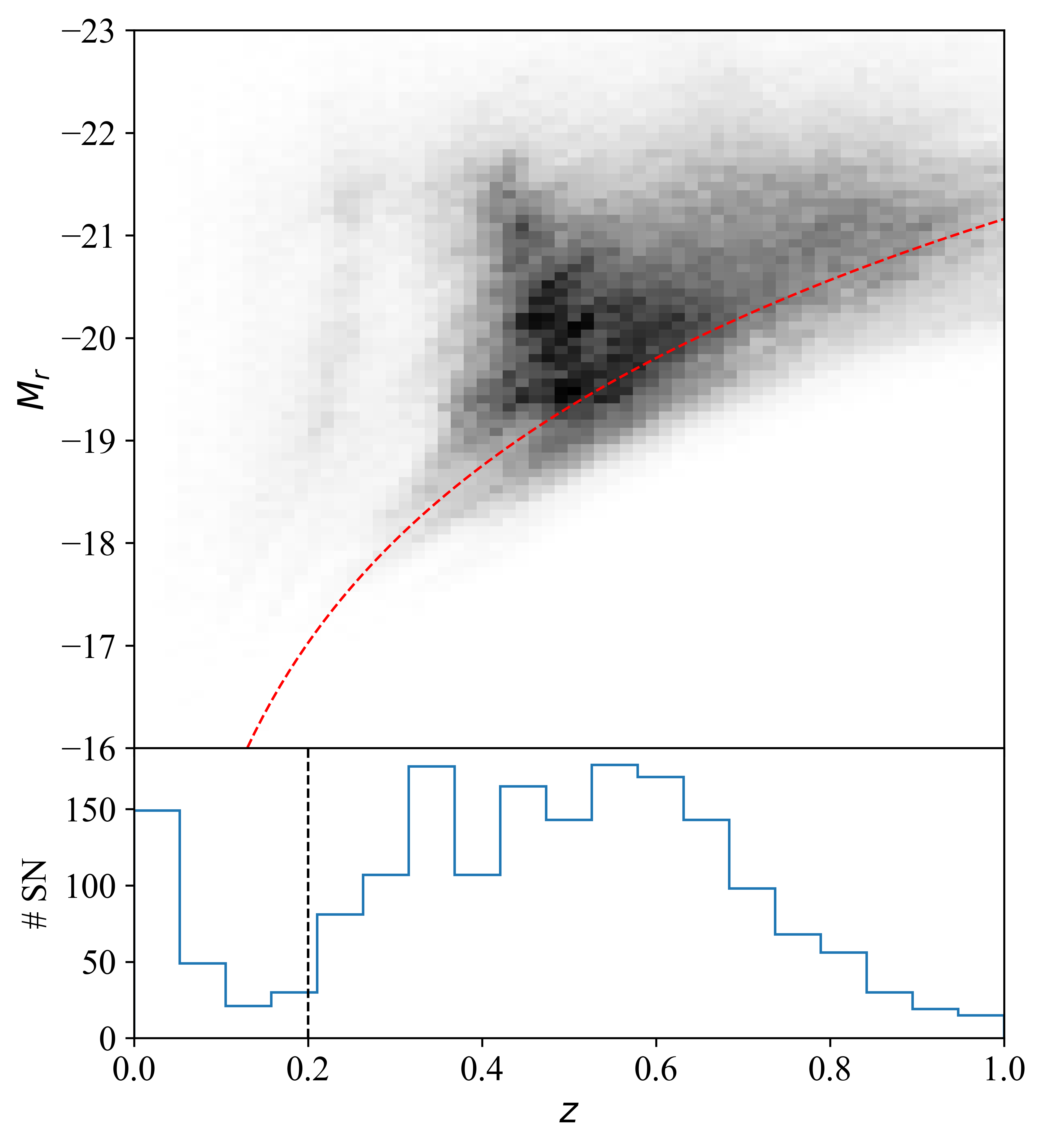}
    \caption{Number distributions of our galaxy (top panel) and supernova (bottom panel) samples by calculated $r$-band absolute magnitude $M_r$ and redshift $z$. The red dotted line shows the Y3 Gold 90\% extended object completeness level of $m_r = 23.0$, and the black dashed line the source redshift cut we use to calibrate our lensing estimator.}
    \label{fig:galaxyandsndist}
\end{figure}

 
Our selected sample therefore comprises 1,503 SN with an average redshift $z \sim 0.53$ and 804,484 galaxies of average redshift $z \sim 0.44$ and is illustrated in Figure \ref{fig:galaxyandsndist}.

\par

\section{Results}
\label{sec:results}

\subsection{Description of the lensing signal}
\begin{figure}
    \centering
    \includegraphics[width=\columnwidth]{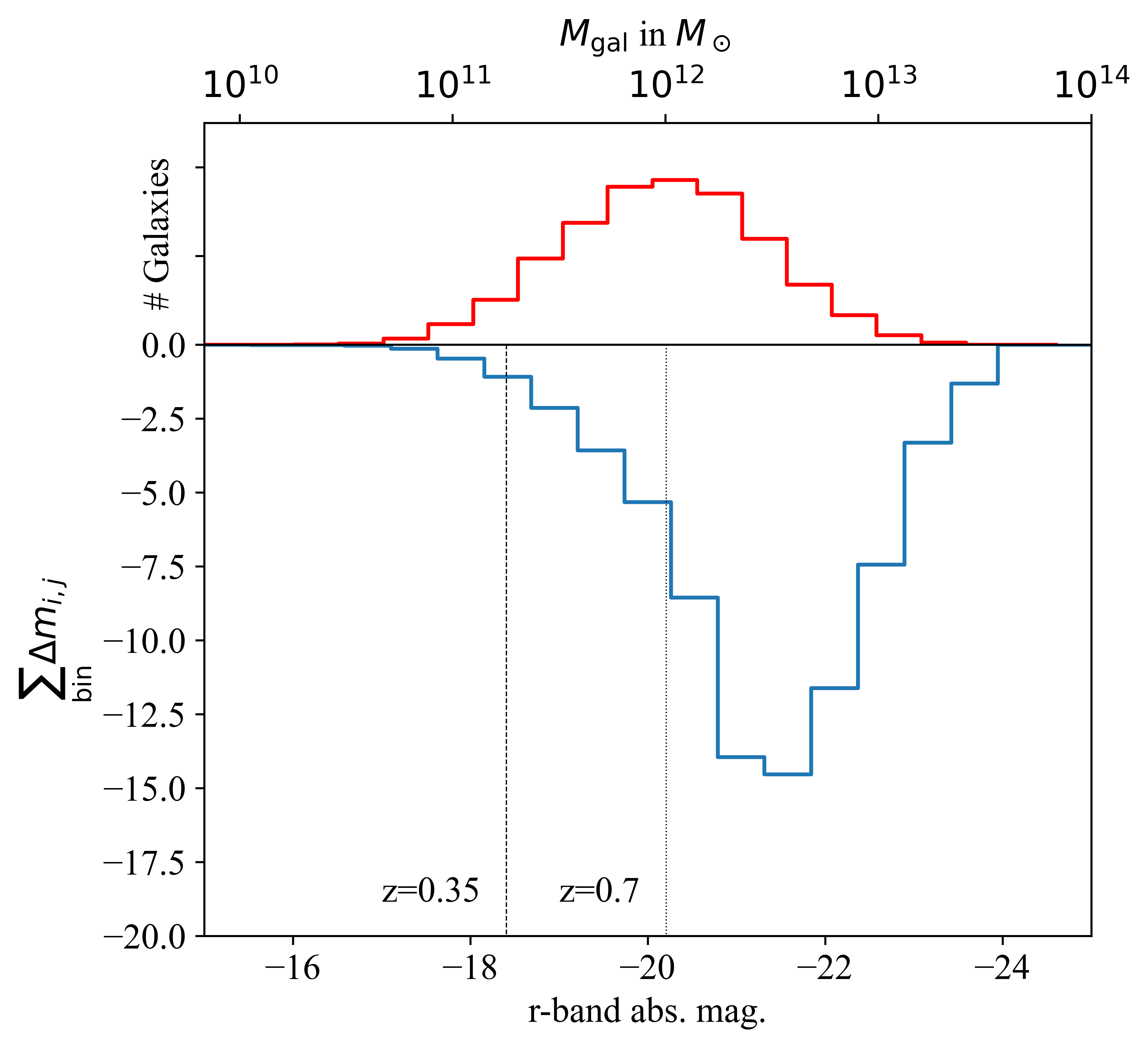}
    \caption{The upper panel shows the counts of galaxies in our entire foreground sample (that is, within the cone delineated by $8\arcmin$ around each SN Ia and bounded by the redshift of the SN Ia) binned by absolute magnitude $M_r$, with $M_{200}$ (using our best fit parameters) shown on the upper x-axis in units of $M_{\odot}$. The lower panel shows our total lensing signal summed over galaxies and binned by absolute magnitude of the galaxy lens. We have marked the Y3 Gold 90\% completeness limit $m_r = 23.0$ for lens redshift $z = 0.35, 0.7$ as the vertical black dashed and dotted lines. The peak of the blue histogram compared to the red shows the majority of our lensing signal is due to foregrounds within the completeness limit of the Y3 Gold catalog.}
    \label{fig:magabsmag}
\end{figure}

The majority of the lensing signal comes from galaxies with impact parameters $b < 300 $ kpc. Although the numbers of galaxies peak at $M_r \sim -20$, the lensing estimate comes predominantly from galaxies with $M_r \sim -21.5$, equivalent to a Milky Way-type galaxy. We illustrate this point in Figure \ref{fig:magabsmag}, where the total lensing estimate calculated for our entire foreground galaxy population is binned by galaxy absolute magnitude. In the plot, we have marked the absolute magnitude of an $m_r = 23.0$ galaxy located at $z = 0.35, 0.7$ to illustrate how the completeness of the foregrounds may affect our lensing signal. 

\par
\begin{figure}
    \centering
    \includegraphics[width=1.0\columnwidth]{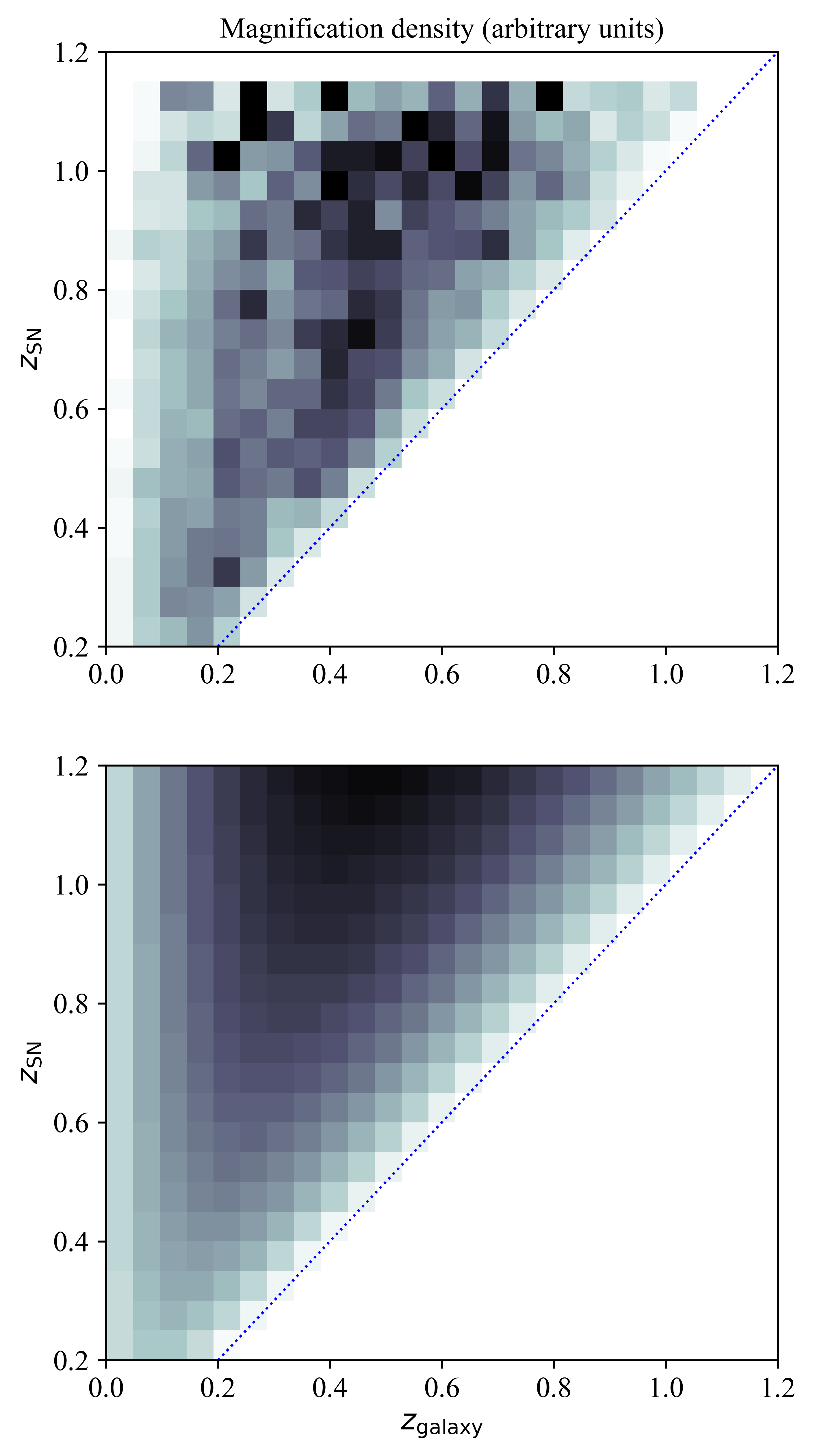}
    \caption{An illustration of the density of the lensing signal per SN Ia as a function of source redshift $z_{\rm SN}$ and lens redshift $z_{\rm galaxy}$. Units are arbitrary and a larger lensing density is represented as a darker box. \textit{Upper panel. } For our data, the lensing peaks as expected, roughly midway between source and observer. The claim that our foregrounds are volume-limited is further supported by continuation of the signal towards high source and foreground redshifts (top right of the figure). \textit{Lower panel. } A theoretical expectation of the dispersion of lensing $\sigma_{\rm lens}$ contributed by lenses in individual redshift bins. This has been computed using the power spectrum model \texttt{HMCODE2020} \citep{Mead2021} and Eqn. (5) of \citet{Frieman1996}. It is apparent that our data conforms to the theoretical expectation, albeit with a high stochasticity.}
    \label{fig:magbysnredshift}
\end{figure}

In Figure \ref{fig:magbysnredshift}, we show an illustration by redshift of where the lensing signal arises for our sample, together with a theoretical expectation derived from an integral over the power spectrum. As expected, it is generally midway in distance between the SN Ia and $z=0$. For a SN Ia at $z \sim 0.7$ and a typical lensing galaxy at redshift $z \sim 0.35$, the completeness limit $m_r = 23.0$ corresponds to $M_r = -18.4$, equivalent to the Large Magellanic Cloud. A slight apparent underdensity in the top right of Figure \ref{fig:magbysnredshift} is due to this limit. As the haloes of the unseen galaxies there still contribute to the true magnification, this can be expected to degrade our correlation result for high redshifts. However, the mass associated with them  will be absorbed on average into the parameter $\Gamma$.  In Section \ref{sec:haloparam} below we estimate how much the limit biases the mass-to-light ratio $\Gamma$.

We inspected the data and images for all fields with $\delta m < -0.1$ to check the reliability of our foreground selection criteria. For SN 1337541 the (spectroscopic) redshift is $z = 1.05$ and the closest catalog galaxy is within $0.5\arcsec$, but has photo-$z$ $\sim 0.3$. Given the discrepancy in the redshifts, we would classify this galaxy as a foreground and not the host. However, it seems probable that the photo-$z$ is contaminated by the light from a larger nearby foreground galaxy and is therefore unreliable. We exclude this SN Ia; this lowers the significance of our results and is therefore conservative. 

\subsection{Halo parameters}
\label{sec:haloparam}
We find $\beta = 2.15 \pm 0.24$ and $\Gamma = 132^{+26}_{-29} \, h \, M_{\odot}/L_{r, \odot}$ where $68\%$ confidence intervals are indicated, and we have used the M08 concentration model. These are population averages for galaxies in the DES Y3 Gold sample, and the error is a combination of statistical uncertainty (such as observational errors in photometry) and natural variation within the confines of our model. This is our fiducial choice of analysis parameters and is used for the figures in this paper. 
\par
\begin{figure}
    \centering
    \includegraphics[width=\columnwidth]{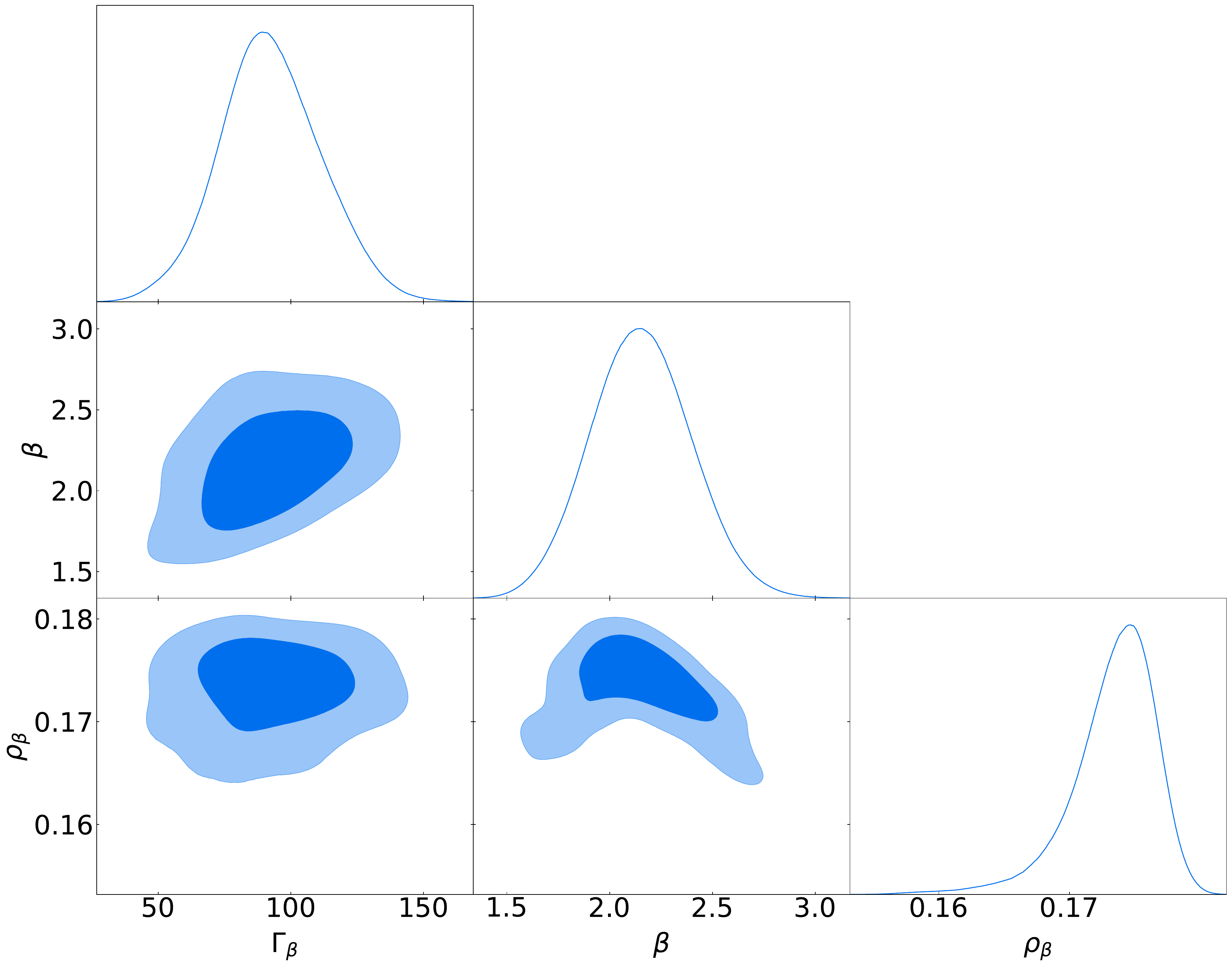}
    \caption{The marginalized posteriors for our power-law halo profile slope, $\beta$, and effective mass-to-light ratio, $\Gamma$. $\beta = 2$ corresponds to the NFW profile. Values for the $\Gamma$ axis are normalised using $h=0.674$.}
    \label{fig:triangleplot}
\end{figure}

Our fiducial result is consistent with an NFW profile $\beta=2$ and also consistent with that obtained from the alternative halo concentration models D08 and C11 to within $1\sigma$. Additionally, letting the concentration vary (as a fixed value), we find $c = 8.3 \pm 3.2$ which is consistent with the average value of $c \sim 6$ from the M08 model. The maximum likelihood values are $\beta = 2.10$ and $\Gamma = 139 \, h \, M_{\odot}/L_{r, \odot}$. The posterior distributions are shown in Figure \ref{fig:triangleplot}.
\par
\begin{figure}
    \centering
    \includegraphics[width=\columnwidth]{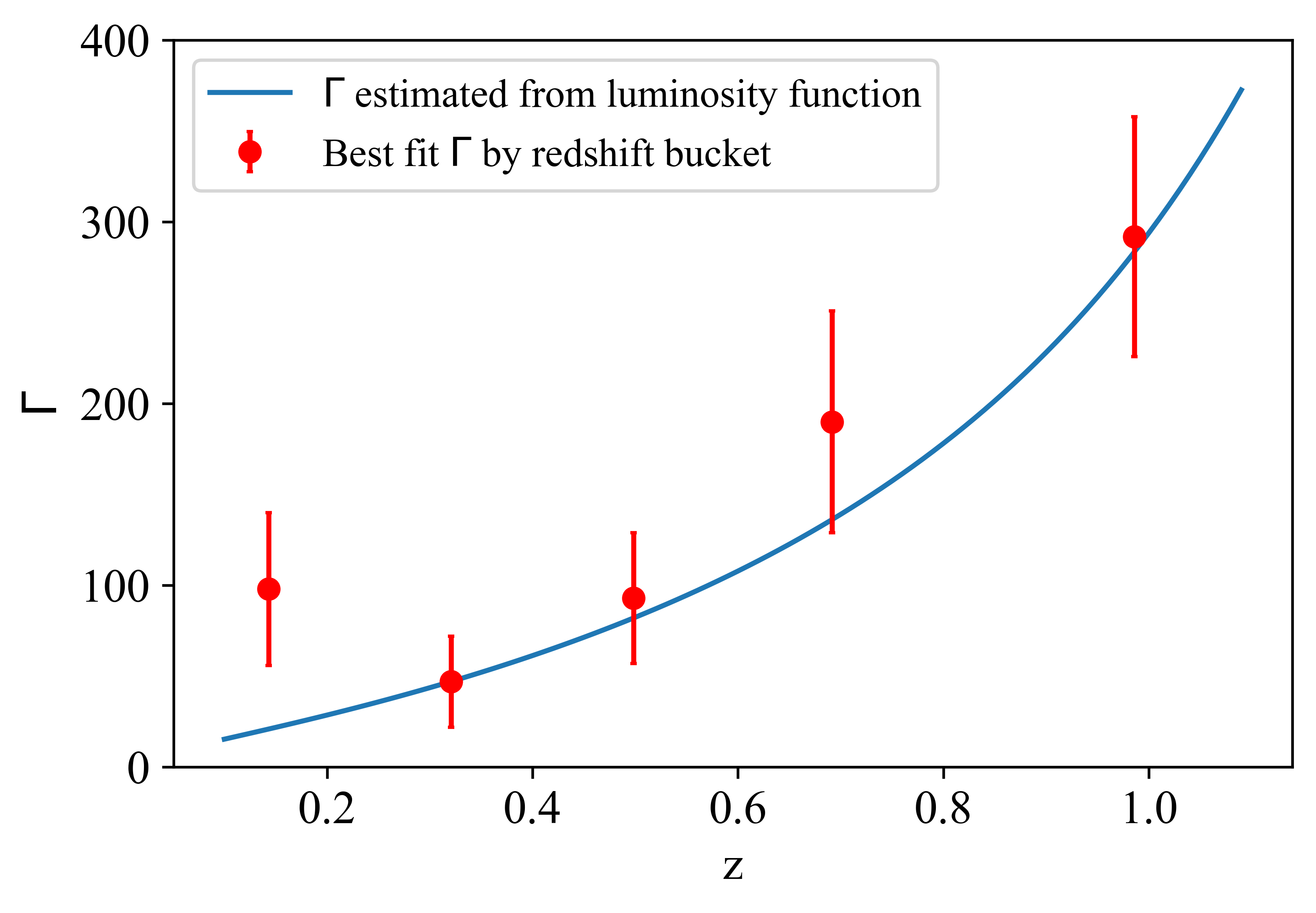}
    \caption{The mass-to-light ratio $\Gamma$ is expected to increase with redshift of the foreground as the Y3 Gold multi-epoch limit at $r \sim 23.6$ decreases the observed light per unit mass for increased distance. The plot compares the trend (for plotting purposes this is normalised at redshift $z=0.32$) from our data with that expected from the galaxy luminosity functions calibrated in \citet{Loveday2011}.}
    \label{fig:gammaredshift}
\end{figure}

 
We also tested the impact of allowing $\Gamma$ to vary with redshift, in broad bins of width $\Delta z=0.2$. As expected $\Gamma$ increases with redshift : distant galaxies are less likely to be in the catalog, but $\Gamma$ must still account for the relation between the magnitude-limited foregrounds and the true physical mass distribution that is lensing. Using the galaxy luminosity functions calibrated in \citet{Loveday2011}, and the Y3 Gold multi-epoch limit of $r = 23.6$, we confirmed that the increase was consistent with expectations from the faint end of the luminosity function, albeit within fairly large error bands. This consistency increases our confidence that our lensing model captures the correct relationship between light and mass, and we plot the results in Figure \ref{fig:gammaredshift}. Alternatively, allowing $\Gamma$ to vary with galactic absolute magnitude indicated a moderate trend to lower values for brighter galaxies, but at no great significance.

\par 
We may compute the fraction of matter bound into virial haloes by summing the implied virial masses of foreground haloes and dividing by the comoving volume enclosed by the cone of radius $8\arcmin$ around the LOS. We find that $\rho_{\rm uniform} = (0.62 \pm 0.11)\rho_m$, in other words that $\sim 40\%$ of matter is bound into haloes. Although this result appears to be consistent with N-body simulations, we note that simulation results are highly dependent on the resolution, and the fraction of matter bound into haloes remains an unsolved problem in cosmology (see discussion in Section 5.1 of \citet{Asgari2023}). It will be particularily interesting to revisit this constraint with future data sets.

\subsection{Correlation of lensing and Hubble diagram residuals}
\label{subsec:correlation}
\begin{figure}
    \centering
    \includegraphics[width=\columnwidth]{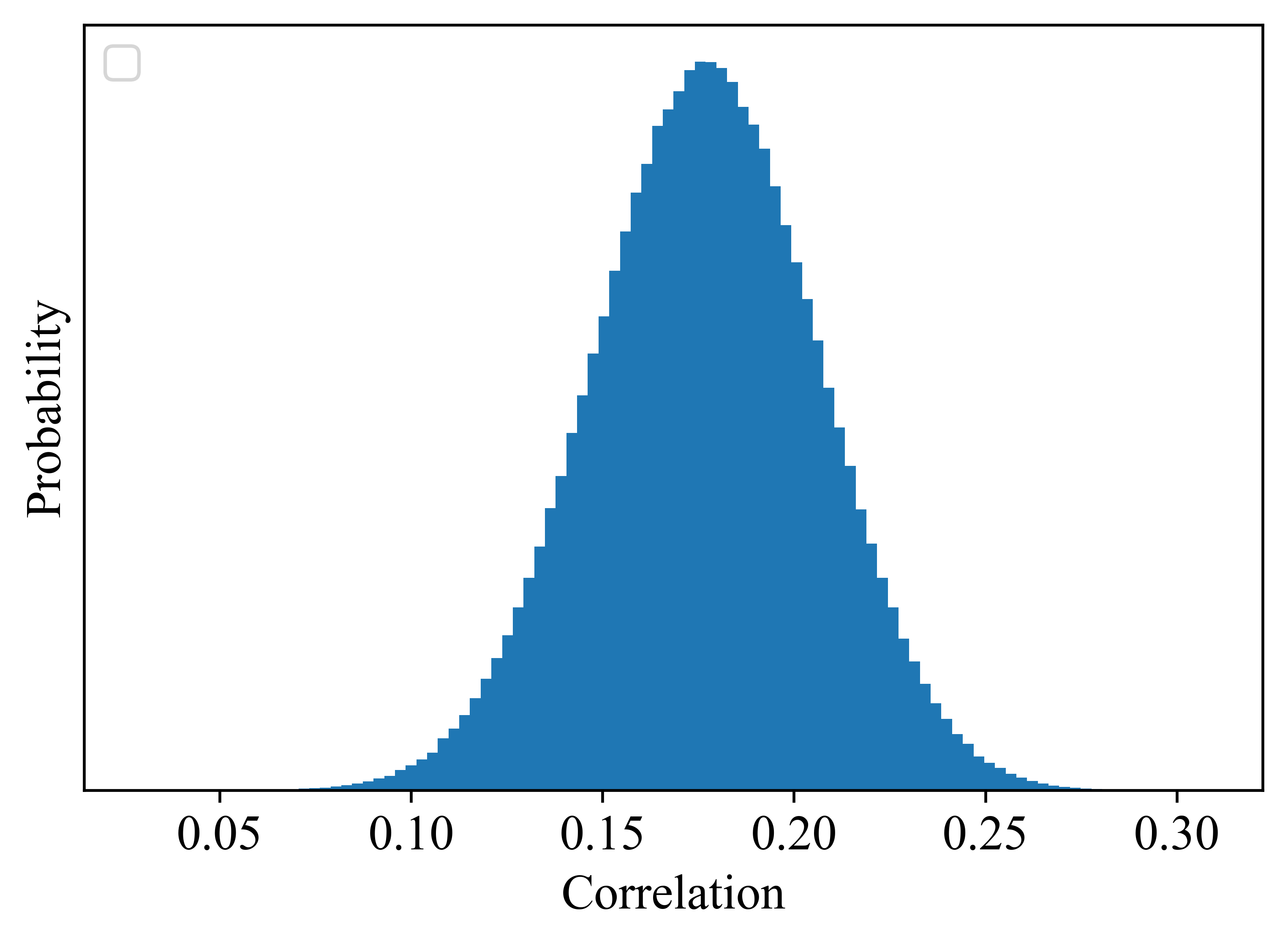}
    \caption{The bootstrap resampling distribution of correlation between our lensing estimator and the Hubble diagram residual for SN Ia of $z>0.2$. The statistical significance of lensing signal detection obtained is $\bar{\rho} / \sigma_{\rho}  = 6.0$. }
    \label{fig:corrbootstap}
\end{figure}

Marginalising over our model parameters, we find a correlation between our lensing estimate $\Delta m$ and Hubble diagram residual $\mu_{\rm res}$ of $\rho = 0.173 \pm 0.029$(stat) for SN Ia with $z>0.2$. The statistical error is derived from $10^6$ bootstrap re-samples of the data as shown in Fig. \ref{fig:corrbootstap}, and corresponds to a significance of $6.0\sigma$ before allowance for systematics. This significance is marginally increased if we allow for a varying $\Gamma$ with redshift as described in the previous section. As seen in Figure \ref{fig:triangleplot}, the correlation is highest close to the mean of $\beta$ and drops outside of our confidence intervals as expected. 

\par 
We test the robustness of our results to our parameter choices, including the thresholds for distinguishing between foreground and hosts, concentration models, and aperture radius. Adopting the standard deviation across our choices as an estimate of potential systematics, we find $\sigma_{\rho} = 0.009$(sys). We conclude that systematics do not materially affect the significance of our correlation. We also checked that the correlation from our pipeline after randomly shuffling the SN Ia residuals was consistent with zero.
\par 
Analysis of the lensing of quasars by foreground galaxies has suggested that approximately a third of lensing magnification may be offset by dust extinction from the foreground galaxies \citep{Menard2010}. However, the effect on the colour parameter $c$ of SN Ia is then smaller than the magnification by a factor of $\sim 10$ (assuming a typical extinction law). This implies that it will not be detectable with our current data set, and indeed we find the correlation between $c$ and $\Delta m$ to be $\rho_{\Delta m, c} = 0.001 \pm 0.026$. We also checked for correlation of our lensing estimator with the stretch parameter $x_1$ and found $\rho_{\Delta m, x1} = 0.040 \pm 0.024$, again consistent with zero.

\begin{figure}
    \centering
    \includegraphics[width=\columnwidth]{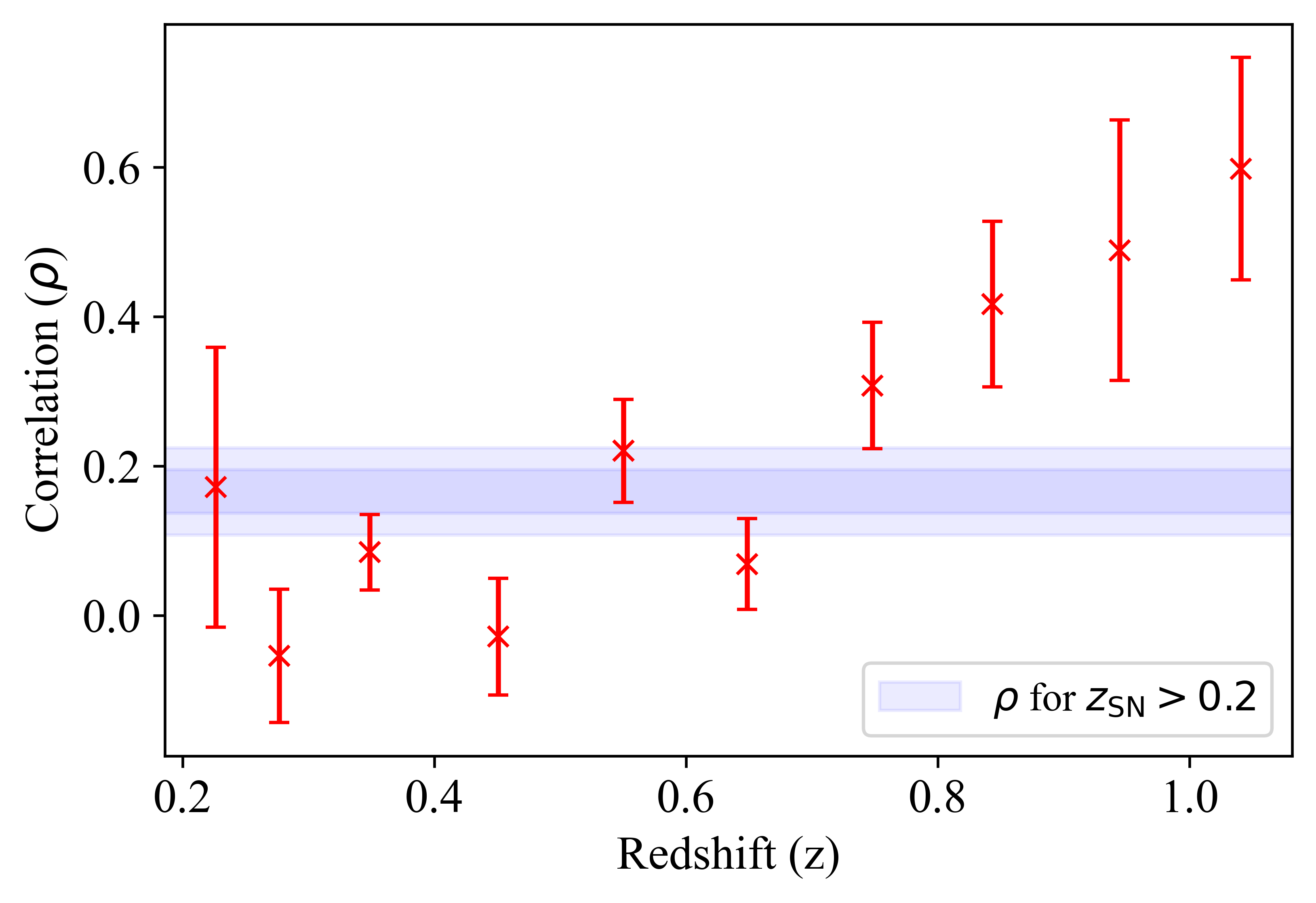}
    \caption{The correlation $\rho$ between the Hubble diagram residuals and weak lensing convergence estimate of our SN Ia sample, shown for individual redshift bins. Errors are computed by bootstrap resampling. As expected for a signal due to lensing, we see a generally increasing trend with distance. The low value in the redshift bin $0.6<z<0.7$ is likely to be a statistical dispersion around the trend; see Fig. \ref{fig:magscatter}.  The horizontal axis shows the average redshift in each bin. Our result of $\rho = 0.173 \pm 0.029$ for the sample between $0.2 < z< 1.2$ is shown as the shaded purple bars at $1\sigma$ and $2\sigma$ confidence. }
    \label{fig:corrbucket}
\end{figure}
\par

\par
\begin{figure*}
    \centering
    \includegraphics[width=17cm]{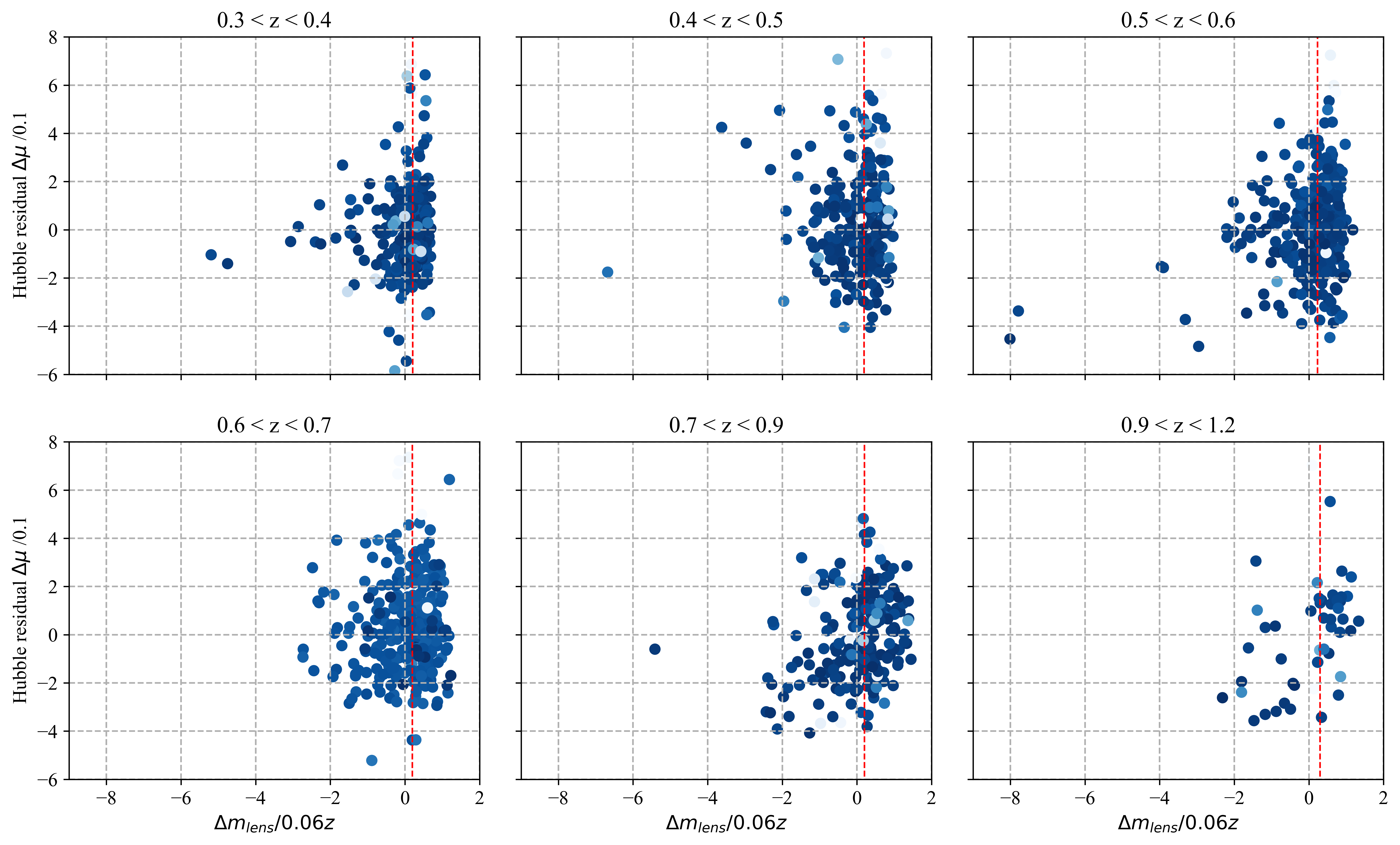}
    \caption{Scatter plots of Hubble diagram residuals $\mu_{\rm res} = \mu - \mu_{\rm model}$ of SN Ia (y-axis) and the lensing estimate $\Delta m$ (x-axis). We have normalised the scales by dividing by the expected lensing dispersion $\sigma_{\rm lens}  = 0.06z$ and intrinsic dispersion $\sigma_{\rm int} = 0.1$. The points are shaded according to the probability they are SN Ia, with lighter blue indicating probable contaminants. The median in each bin is marked with a dashed red line. For low redshift bins, the scatter plots are dominated by the intrinsic dispersion of magnitudes with little visible correlation with the lensing estimate. For higher redshift bins, the correlation is apparent as the clustering of points in the top right quadrant (the majority of lines of sight are through underdense regions) and a small number of magnified supernovae in the bottom left. }
    \label{fig:magscatter}
\end{figure*}

In Figure \ref{fig:corrbucket} we show the correlation per redshift bin. As expected, the correlation shows an increasing trend with distance as the lensing becomes a greater proportion of the Hubble diagram residual scatter. We show scatter plots of our residuals in Figure \ref{fig:magscatter}. The median of the lensing estimator in each bucket is marked with a red dashed line. The median is greater than the (zero) mean, showing the majority of SNe Ia are de-magnified and a smaller number of SNe Ia are magnified. The intrinsic scatter dominates for low redshifts, but for larger redshifts the correlation is visible as the grouping of dots towards the bottom left and top right quadrants. 

\subsection{Lensing dispersion}
\label{sec:lensdispersion}
As noted in S22, from general principles we expect $\sigma_{\rm lens} \propto d_M(z_s)^{3/2}$  where $d_M(z_s)$ is the comoving distance to a source at redshift $z_s$.\footnote{It is common in the literature \citep[for example, see][]{Jonsson2010, Holz2005} to linearise this such that $\sigma_{\rm lens} \propto z_s$.} This was derived in S22 on the assumption that the mass function and comoving number density of haloes is constant over the redshift range of our galaxy sample. 
\par
\begin{figure}
    \centering
    \includegraphics[width=\columnwidth]{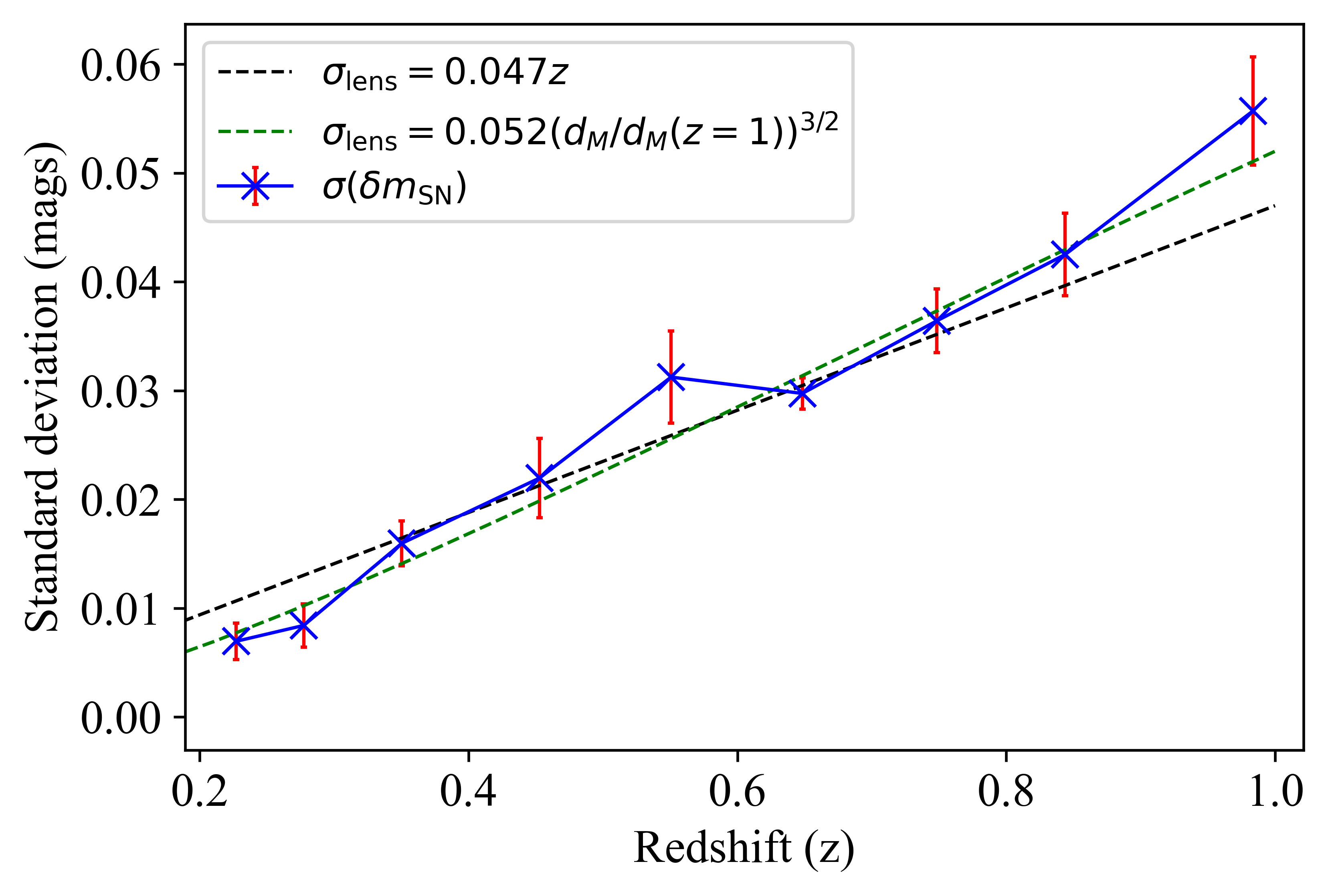}
    \caption{The standard deviation of $\Delta m_{\rm lens}$ as computed from the actual lines of sight to the DES 5Y SN Ia sample.}
    \label{fig:lensdisperse}
\end{figure}

Considering the dispersion of our lensing estimator between individual SN Ia in a given redshift bucket, we can fit for $\sigma_{\rm lens}(z) = A \times d_{\rm M}(z)^{B}$ where $A,B$ are constants. We find $B = 1.55 \pm 0.12$, which is consistent with expectations. Accordingly, we fix $B=1.5$ and we then find 
\begin{equation}
\label{eq:lensvar}
    \sigma_{\rm lens} = (0.052 \pm 0.009) (d_{\rm M}(z)/ d_{\rm M}(z=1))^{3/2} \;\;,
\end{equation}
where the fit is shown in Figure \ref{fig:lensdisperse}. We have normalized the above using $d_{\rm M}(z=1)$ to facilitate comparison with the literature. Our result is consistent within errors for $z \le 1$ of the commonly cited $\sigma_{\rm lens} = 0.055z$ \citep{Jonsson2010}, but discrepant with $\sigma_{\rm lens} = 0.088z$ \citep{Holz2005} at $>3\sigma$. We note that \citet{Holz2005} was derived from simulations which added additional lensing due to compact objects. This suggests the DES-SN5YR dataset may be used to place limits on the presence of compact objects, and this will be explored in a future paper.
\par
As the intrinsic SN Ia scatter is $\sigma_{\rm int} \sim 0.1$ \citep{Brout2019}, the dispersion in magnitude caused by lensing will be comparable to it by $z \sim 2$. 
\par
We also calculated the dispersion from two samples of 10,000 random LOS using our best fit halo model parameters. The first sample was generated by allocating LOS to random SN Ia hosts in DES footprint according to the observed SN Ia redshift distribution. The second sample was a random selection of sky positions; these are very unlikely to be near a putative host galaxy. The dispersion of lensing estimator for the former (random host SN Ia) was consistent with the dispersion of the DES-SN5YR Ia sample. This demonstrates that the DES-SN5YR sample is large enough to represent the pdf and obtain an observational lensing dispersion. Interestingly, the dispersion for the latter (random position SN Ia) LOS was larger for redshift $z>0.5$, with $\sigma_{\rm lens} = 0.083$ at $z \sim 1$. This discrepancy was also noted in \citet{Jonsson2010}, who compared the SNLS sample to a random one. This may suggest factors (such as obscuration of distant galaxies by crowded foregrounds) that have biased the observation of SN Ia to lines of sight with a \textit{lower} matter inhomogeneity. 
\par

\subsection{Delensing SN Ia}
We expect that subtracting the lensing estimate will reduce the residuals to the Hubble diagram. We therefore propose a modification of the Tripp estimator as 
\begin{equation}
\label{eq:trippnew}
\mu_{\rm delens} = m_B - M_B + \alpha x_1 - \beta c  + \Delta_{\rm M} + \Delta_{\rm B} - \eta \Delta m_{\rm lens}\; .
\end{equation}
In this equation, $m_B, x_1$ and $c$ are parameters that are fitted to the SN Ia light curves representing the amplitude, duration and colour respectively of the observations. $\Delta_{\rm M}$ is an adjustment to take account of variations in SN Ia magnitudes correlated to their host galaxy properties (usually summarized by host stellar mass $M_*$), and $\Delta_{\rm B}$ is a term to correct for Malmquist bias and computed from simulations. The novel term we propose is the last term, $\eta \Delta m_{\rm lens}$ which is calculated using best-fit model parameters for each individual SN Ia. In this context, lensing becomes simply a second environmental variable equivalent to (and of similar size as) the host mass step adjustment $\Delta_{\rm M}$, and the distance moduli $\mu$ are \say{de-lensed}. 
\par
To fit cosmological parameters, we will use Eqn. \ref{eq:chi2} with the covariance $C$ reduced as per Eqns. \ref{eq:like1}, \ref{eq:like2} and with $\mu$ replaced by $\mu_{\rm delens}$.
\par
A version of Eqn. \ref{eq:trippnew} was proposed in \citet{Smith2014}, where an estimator was constructed from the local number density of a spectroscopic sample. However the density of the spectroscopic sample will vary over the survey footprint, necessitating a spatial calibration of $\eta$. This is less practical than our model,  as our halo parameters are already calibrated to the average relationship between our foreground tracer and mass. Consequently, we expect (and recover, see below) $ \eta \sim 1$ but the addition of this free parameter provides a convenient cross-check on the maximum likelihood values for $\Gamma, \beta$ used to construct $\Delta m_{\rm lens}$.
\par
In the original analysis without our new term, lensing effects have been incorporated in two places. Firstly, the covariance matrix has had an additional noise of $\sigma_{\rm lens} = 0.055z$ added to the diagonal; we have corrected this as noted above. Secondly, a more subtle issue is that $\Delta_{\rm B}$, which is calculated by the code package \texttt{SNANA}\footnote{https://github.com/RickKessler/SNANA} \citep{Kessler2009}, incorporates (amongst other effects) a redshift-dependent Malmquist bias correction derived from lensing pdfs from N-body simulations. 
\par 
These pdfs under-estimate $\sigma_{\rm lens}$ compared to Eqn. \ref{eq:lensvar} by about $30\%$. There are two potential solutions. Firstly, we may re-calculate the bias calculation by either scaling the existing pdfs to match our observed $\sigma_{\rm lens}$, or by generating new pdfs using the code package \texttt{TurboGL}\footnote{https://github.com/valerio-marra/turboGL} \citep{Kainulainen2009} which uses a similar density model to our method (and would -- correctly -- introduce a dependency of the bias correction on $\sigma_8$). Alternatively, a consistent approach would be to recalculate the bias corrections using our modified Tripp estimator, together with foregrounds simulated to match the distribution of observations. In this case $\eta$ would be treated as a free nuisance parameter on the same footing as $\alpha$ and $\beta$ in Eqn. \ref{eq:trippnew}. For the purposes of this paper, we assume changes to the $\Delta_{\rm B}$ represent second order adjustments to our results, as the current input model is not too far from values derived from the data.
\par
\begin{figure}
    \centering
    \includegraphics[width=\columnwidth]{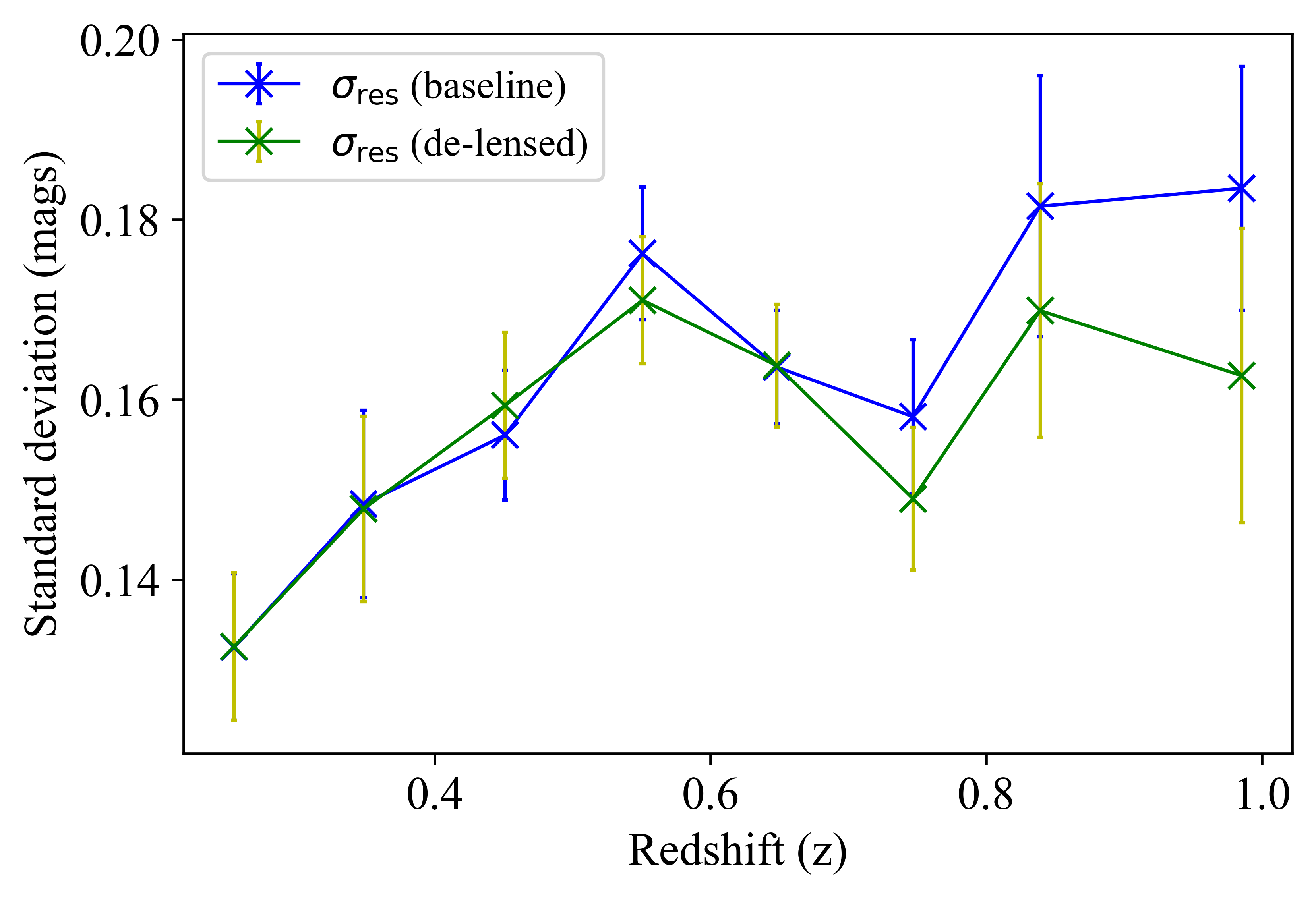}
    \caption{The standard deviation of Hubble diagram residuals for de-lensed SN Ia (green) and the original residuals with lensing dispersion (blue). For illustrative purposes, we have removed potential contaminants and less well-observed SN Ia with $\sigma_{\mu} > 0.25$. Thus for this high-purity sample, we see that the dispersion of the de-lensed residuals has a reduced upwards trend compared to the baseline data.}
    \label{fig:delensresid}
\end{figure}

 
Figure \ref{fig:delensresid} shows the delensed residuals constructed using Eqn. \ref{eq:trippnew}, with the maximum likelihood model parameters given in Section \ref{sec:haloparam}. For the purposes of the figure, we have selected a \say{high purity} sample with statistical error $\sigma_{\mu} < 0.25$ (otherwise the errors would be dominated by likely non-SN Ia contaminants). While the residual scatter increases with redshift remains, de-lensing has reduced the trend. In particular, it is remarkable that the de-lensed residuals for SNe Ia with $0.9<z<1.0$ exhibit no more scatter than those in the $0.4<z<0.5$ bucket. 
\par
Eqn. \ref{eq:trippnew} may be used to re-compute cosmological parameters. For cosmological parameter baseline, we use the entire SN Ia dataset and likelihood as described in \citet{SNKeyPaper}. For the delensed inference, we use the delensed distance moduli $\mu_{\rm delens}$ from Eqn. \ref{eq:trippnew} with $\eta = 1$ and set $\Delta m_{\rm lens} =0$ for $z<0.2$, replacing the covariance with adjusted matrix given in Eqns. \ref{eq:like1}, \ref{eq:like2}. We have tested our results are consistent if we marginalise over $\eta$ as a free parameter. Fitting is done in Polychord, with flat priors $\Omega_{\rm M} \in (0.1,0.5)$ and $w \in (-1.5,-0.5)$.
\par
Our results are shown in Table \ref{tab:cosmo}. Our baseline values are consistent with those reported in \citet{SNKeyPaper}. In Flat-$\Lambda$CDM, we find $\Delta \Omega_{\rm M} = +0.005$, or about $0.3\sigma$. For Flat-$w$CDM, we find $\Delta \Omega_{\rm M} = +0.036$ and $\Delta w = -0.056$, again about $0.3\sigma$ shift in parameters. In terms of the deceleration parameter $q_0 = \ddot{a}a/\dot{a}^2(z=0)$, we find a change of $-0.017$ to $q_0 = -0.402$. 
\par
De-lensing thus moves Flat $w$CDM parameters somewhat closer to Flat $\Lambda$CDM, and suggests the observed SN Ia are on slightly under-dense LOS. Reassuringly, the change in cosmological parameters by correcting for lensing is not large in DES-SN5YR, even though line-of-sight biases may still have arisen in the spectroscopic confirmation of the host redshift. It is possible that for future datasets probing SN Ia at higher redshift, obscuration by foregrounds may also introduce line-of-sight bias if a delensing term is not used. 

\renewcommand{\arraystretch}{1.5}
\begin{table}
\label{tab:cosmo}
\centering
\begin{tabular}{l l l l} 
 \hline
  & $\Omega_{\rm M}$ & $w$ & $\chi^2$ \\
 \hline 
 Flat-$\Lambda$CDM \\
 \hline
 Baseline & $0.354 \pm 0.016$ & - & 1640 \\
 Delensed & $0.359 \pm 0.016$ & - & 1646 \\
 \hline
 Flat-$w$CDM \\
 \hline
 Baseline & $0.258 ^{+0.095}_{-0.070}$ & $-0.81 ^{+0.17}_{-0.13}$ & 1638 \\
 Delensed & $0.294 ^{+0.087}_{-0.062}$ & $-0.87 ^{+0.18}_{-0.14}$ & 1646 \\
 \hline
\end{tabular}
\caption{Marginalised mean values and $68\%$ confidence intervals for cosmological parameters before and after delensing. Note that the $\chi^2$ values here should \textit{not} be interpreted as a relative goodness-of-fit, as the covariance matrix for the delensed case has been adjusted to remove the original noise term $0.055z$ allocated to lensing. Keeping the covariance matrix unchanged results in a $\Delta \chi^2 \sim -55$ preference for the delensed model. The consistency of the $\chi^2$ between the two models shows delensing is effective at removing the majority of the previously-assumed noise. }
\end{table}

\subsection{Constraints on inhomogeneity}
\label{sec:inhomogeneity}

A theoretical prediction for $\sigma_{\rm lens}$ may be made from an integral over the matter power spectrum and redshift \citep{Frieman1996}, together with a prefactor proportional to the physical matter density $\Omega_{\rm M}h^2$. 
\par
This may be taken to imply that an observed value for $\sigma_{\rm lens}$ may then be used to constrain the amplitude of the power spectrum, or equivalently $\sigma_8$. However, there are many theoretical and observational issues to overcome. We have earlier noted that the dispersion of our sample may be suppressed due to extinction and obscuration by foregrounds. Also, the sensitivity of the integrand extends well into the non-linear regime $k>1$ Mpc$^{-1}$, meaning both baryonic feedback and the presence (or not) of compact objects would alter the theory expectation. Values from ray-tracing in N-body simulations are therefore likely to be sensitive to the particle mass and gravity softening scale. These effects may all be of similar size and conspire to offset.
\par 
Ignoring these objections for now, in \citet{Marra2013} the \texttt{TurboGL}\footnote{https://github.com/valerio-marra/turboGL} simulation code \citep{Kainulainen2009} was used to construct a fitting formula for $\sigma_8(\sigma_{\rm lens}(z), \Omega_{\rm M})$. \texttt{TurboGL} simulates weak lensing by randomly placing smooth NFW-profile dark matter halos along the line of sight, from which the magnification due to each halo is calculated semi-analytically. Many such simulations are run to assemble a lensing magnification pdf. The halo masses and number counts are drawn from literature halo mass functions, from which arise the dependence on $\sigma_8$ and $\Omega_{\rm M}$.
\par
Using the fitting formula from Equation 6 of \citet{Marra2013}, with priors of $\Omega_{\rm M} = 0.315 \pm 0.007$ and $H_0 = 67.4 \pm 0.5$ (see Table 2 of \citet{SNKeyPaper}, these are from a combined analysis incorporating likelihoods from the CMB \citep{Planck2018} and DES 3x2pt weak lensing results \citep{Abbott2022}), the dispersion of our lensing estimator calibrated to the DES Y5 SN Ia sample gives
\begin{equation}
    \sigma_8 = 0.90 \pm 0.13 \;\; .
\end{equation}
Given the larger error bars, this is consistent both with results from the DES 3x2pt analysis \citep{Abbott2022} and from Planck \citep{Planck2018}. However, we caution the reader that this consistency may be largely coincidental for the reasons discussed above.

\section{Summary and discussion}
\label{sec:summary}
In this paper, we have forward-modelled the weak lensing convergence for individual SNe Ia based on the astrometric and photometric properties of foreground galaxies with two free model parameters. We have demonstrated that the assumptions of our model form an effective statistical basis for constructing an estimator that correlates significantly with SN Ia residuals to their Hubble diagram. We find $\rho  =0.177 \pm 0.029$, a detection of non-zero correlation at $6.0\sigma$ significance.
\par
Our results are consistent with expectations from the literature. \citet{Kronborg2010} detected the presence of lensing at $2.3\sigma$ significance using a sample of 171 SN Ia selected from the supernova legacy survey (SNLS), with certain assumptions about the profile of dark matter haloes and relationship between mass and light. \citet{Jonsson2010} found a significance of $1.4\sigma$ with a similar sample, but relaxing some of those assumptions. \citet{Smith2014} found a significance of $1.4\sigma$ using a sample of 749 SN Ia from the Sloan Digital Sky Survey (SDSS) and an estimator was based on number counts spectroscopically measured foregrounds.
\par 
\citet{Kronborg2010} forecast a $3\sigma$ detection with a sample of 400 SNLS-like SN Ia. Our results are consistent with this forecast; as can be seen from Figure \ref{fig:triangleplot} forcing a non-data driven halo shape, as they do, would lower the measured correlation. While \citet{Smith2014}, used a larger sample of $\sim$800 SN Ia, the SDSS survey is shallower than SNLS and the use of (sparser) spectroscopic-only foregrounds and an estimator based on number counts (somewhat equivalent to forcing $\beta=0$ in our model) will significantly dampen the signal. Our fit for $\sigma_{\rm lens}$ is consistent with \citet{Jonsson2010}. However, it is lower than the prediction of \citet{Holz2005}, due to the fact we do not allow for the (hypothetical) presence of compact objects which increase the dispersion. It is then likely that DES-SN5YR can be used to constrain the number density of compact objects close to the lines of sight, but we leave this to future work.
\par 
In summary, our results pass a $5\sigma$ significance level for the first time in the literature by the use of the larger, deeper DES-SN5YR sample and an optimal estimator.
\par
Confidence in our model is supported by the fact that the model parameter posteriors encompass physically reasonable values. Adjusting for the percentage of foreground galaxies we excluded due to unreliable photo-z estimates, we find that the mass-to-light ratio between DES Y3 Gold r-band catalog magnitudes and virialised halo mass $M_{200}$ is $\Gamma = 143^{+28}_{-32} \, h \, M_{\odot}/L_{r, \odot}$, which is broadly in line with expectations. The mass-to-light ratio increases with lens redshift in a way consistent with expectations from galaxy luminosity functions. The correlation increases with higher redshift buckets as lensing forms an increasing fraction of the observational dispersion of SN Ia magnitudes. We find that the lensing of SN Ia implies that $41\% \pm 12\%$ of matter is bound into virial haloes. 
\par
We have shown that when our estimator is used as an additional variable in the standardization of SN Ia magnitudes, it lowers the scatter of Hubble diagram residuals, again to greater effect in high redshift buckets. When we re-compute cosmological parameters using our delensed distance moduli, the change is small for the DES-SN5YR sample, in the direction corresponding to the dataset having been on slightly under-dense sight lines. As we have calibrated our estimator to DES Y3 Gold photometry, it may in principle be applied to \textit{any} line of sight in that footprint. It therefore may be used to construct maps of the model convergence across the footprint (also known as \say{mass maps}) to complement, or augment, those derived from galaxy-galaxy lensing \citep[for example, as given in][]{Jeffrey2021}. The investigation of this will be left to future work.
\par
Given the modest change in cosmological parameters from de-lensing it may be tempting to conclude it is not particularly relevant to homogeneous cosmological parameters. This would be hasty for two reasons. Firstly, for data extending to higher redshift than DES-SN5YR, it is not guaranteed that de-lensing will continue to be a small change even for photometrically confirmed datasets. In particular, \citet{Weinberg1976} has pointed out increasing obscuration due to foregrounds could bias cosmological parameters to under-dense lines of sight. Thus, we would expect the tests we have proposed in this paper to be relevant to the forthcoming SN Ia survey of the Nancy Grace Roman Space Telescope \citep[]{Hounsell2018}. Secondly, with $>1,000,000$ SN Ia expected from the forthcoming Rubin LSST survey \citep[]{LSST2019}, systematics can be expected to be a limiting factor in determining cosmological parameters. In particular, the portion of the Malmquist bias correction due to lensing will be a large contribution. If lensing effects are better constrained, the systematic uncertainty can be lowered.
\par 
Our results open a new pathway in the use of SN Ia observations to study inhomogeneities. A $\sigma_8$ constraint was derived by comparing the observational dispersion of our lensing estimator to a literature fit from simulations. Additionally, the presence (or not) of compact objects both increases the expected dispersion \citep{Holz2005} and introduces a specific, redshift-dependent skew signal to the Hubble diagram residuals. However, we note that systematics of this procedure remain unexplored at present. Anticipating that they may be controlled in future work, we expect that SN Ia may be used to complement and enhance existing weak lensing results, and investigate the distribution of matter on both linear and non-linear scales.

\section*{Contribution Statement and Acknowledgements}
PS devised the project, compiled the data, performed the analysis and drafted the manuscript; DBa, TD, JF, LG, DH, RK, JL, CL, OL, RM, RN, MSa, MSu, MV, PW advised on the analysis and commented on the manuscript; DBa and JF were also internal reviewers, and RM the final reader. The remaining authors have made contributions to this paper that include, but are not limited to, the construction of DECam and other aspects of collecting the data; data processing and calibration; developing broadly used methods, codes, and simulations; running the pipelines and validation tests; and promoting the science analysis.

This paper has gone through internal review by the DES collaboration. Funding for the DES Projects has been provided by the U.S. Department of Energy, the U.S. National Science Foundation, the Ministry of Science and Education of Spain, the Science and Technology Facilities Council of the United Kingdom, the Higher Education Funding Council for England, the National Center for Supercomputing 
Applications at the University of Illinois at Urbana-Champaign, the Kavli Institute of Cosmological Physics at the University of Chicago, the Center for Cosmology and Astro-Particle Physics at the Ohio State University,
the Mitchell Institute for Fundamental Physics and Astronomy at Texas A\&M University, Financiadora de Estudos e Projetos, Funda{\c c}{\~a}o Carlos Chagas Filho de Amparo {\`a} Pesquisa do Estado do Rio de Janeiro, Conselho Nacional de Desenvolvimento Cient{\'i}fico e Tecnol{\'o}gico and the Minist{\'e}rio da Ci{\^e}ncia, Tecnologia e Inova{\c c}{\~a}o, the Deutsche Forschungsgemeinschaft and the Collaborating Institutions in the Dark Energy Survey.

The Collaborating Institutions are Argonne National Laboratory, the University of California at Santa Cruz, the University of Cambridge, Centro de Investigaciones Energ{\'e}ticas, Medioambientales y Tecnol{\'o}gicas-Madrid, the University of Chicago, University College London, the DES-Brazil Consortium, the University of Edinburgh, the Eidgen{\"o}ssische Technische Hochschule (ETH) Z{\"u}rich, Fermi National Accelerator Laboratory, the University of Illinois at Urbana-Champaign, the Institut de Ci{\`e}ncies de l'Espai (IEEC/CSIC), 
the Institut de F{\'i}sica d'Altes Energies, Lawrence Berkeley National Laboratory, the Ludwig-Maximilians Universit{\"a}t M{\"u}nchen and the associated Excellence Cluster Universe, the University of Michigan, NSF's NOIRLab, the University of Nottingham, The Ohio State University, the University of Pennsylvania, the University of Portsmouth, SLAC National Accelerator Laboratory, Stanford University, the University of Sussex, Texas A\&M University, and the OzDES Membership Consortium.

Based in part on observations at Cerro Tololo Inter-American Observatory at NSF's NOIRLab (NOIRLab Prop. ID 2012B-0001; PI: J. Frieman), which is managed by the Association of Universities for Research in Astronomy (AURA) under a cooperative agreement with the National Science Foundation. Based in part on data acquired at the Anglo-Australian Telescope. We acknowledge the traditional custodians of the land on which the AAT stands, the Gamilaraay people, and pay our respects to elders past and present. Parts of this research were supported by the Australian Research Council, through project numbers CE110001020, FL180100168 and DE230100055. Based in part on observations obtained at the international Gemini Observatory, a program of NSF’s NOIRLab, which is managed by the Association of Universities for Research in Astronomy (AURA) under a cooperative agreement with the National Science Foundation on behalf of the Gemini Observatory partnership: the National Science Foundation (United States), National Research Council (Canada), Agencia Nacional de Investigaci\'{o}n y Desarrollo (Chile), Ministerio de Ciencia, Tecnolog\'{i}a e Innovaci\'{o}n (Argentina), Minist\'{e}rio da Ci\^{e}ncia, Tecnologia, Inova\c{c}\~{o}es e Comunica\c{c}\~{o}es (Brazil), and Korea Astronomy and Space Science Institute (Republic of Korea).  This includes data from programs (GN-2015B-Q-10, GN-2016B-LP-10, GN-2017B-LP-10, GS-2013B-Q-45, GS-2015B-Q-7, GS-2016B-LP-10, GS-2016B-Q-41, and GS-2017B-LP-10; PI Foley).  Some of the data presented herein were obtained at Keck Observatory, which is a private 501(c)3 non-profit organization operated as a scientific partnership among the California Institute of Technology, the University of California, and the National Aeronautics and Space Administration (PIs Foley, Kirshner, and Nugent). The Observatory was made possible by the generous financial support of the W.~M.~Keck Foundation.  This paper includes results based on data gathered with the 6.5 meter Magellan Telescopes located at Las Campanas Observatory, Chile (PI Foley), and the Southern African Large Telescope (SALT).
The authors wish to recognize and acknowledge the very significant cultural role and reverence that the summit of Maunakea has always had within the Native Hawaiian community. We are most fortunate to have the opportunity to conduct observations from this mountain.

The DES data management system is supported by the National Science Foundation under Grant Numbers AST-1138766 and AST-1536171. The DES participants from Spanish institutions are partially supported by MICINN under grants ESP2017-89838, PGC2018-094773, PGC2018-102021, SEV-2016-0588, SEV-2016-0597, and MDM-2015-0509, some of which include ERDF funds from the European Union. IFAE is partially funded by the CERCA program of the Generalitat de Catalunya.
Research leading to these results has received funding from the European Research Council under the European Union's Seventh Framework Program (FP7/2007-2013) including ERC grant agreements 240672, 291329, and 306478. We  acknowledge support from the Brazilian Instituto Nacional de Ci\^encia e Tecnologia (INCT) do e-Universo (CNPq grant 465376/2014-2).

This research used resources of the National Energy Research Scientific Computing Center (NERSC), a U.S. Department of Energy Office of Science User Facility located at Lawrence Berkeley National Laboratory, operated under Contract No. DE-AC02-05CH11231 using NERSC award HEP-ERCAP0023923.

This manuscript has been authored by Fermi Research Alliance, LLC under Contract No. DE-AC02-07CH11359 with the U.S. Department of Energy, Office of Science, Office of High Energy Physics.

\section*{Data Availability}
The data and Python code used to generate the results and plots in this paper are available on reasonable request from the authors. A file containing $\Delta m_{\rm lens}$ to generate de-lensed SN Ia distance moduli (Equation \ref{eq:trippnew}) will be added to the DES-SN5YR data release package upon publication.
 



\bibliographystyle{mnras}
\bibliography{DEScitations} 


\bsp	
\label{lastpage}
\end{document}